\documentclass[aps,prd,showkeys,nofootinbib,superscriptaddress,11pt]{revtex4-2}

\usepackage{color}
\definecolor{darkgreen}{rgb}{0,0.35,0}

\usepackage{hyperref}
\usepackage{amsfonts}
\usepackage{amsmath}
\usepackage{bbold}
\usepackage[noabbrev]{cleveref}
\usepackage{subfig}
\usepackage{mathtools,slashed,tensor}
\usepackage[font=scriptsize,labelfont=bf]{caption}
\usepackage{float}
\usepackage[top=2cm, bottom=2cm, left=3cm, right=3cm]{geometry}
\usepackage{tikz-cd}
\usepackage{siunitx}

\newcommand{\be}{\begin{equation}}

\newcommand{\ee}{\end{equation}}
\newcommand{\bea}{\begin{eqnarray}}
\newcommand{\eea}{\end{eqnarray}}
\newcommand{\bef}{\begin{figure}}
\newcommand{\eef}{\end{figure}}
\newcommand{\bce}{\begin{center}}
\newcommand{\ece}{\end{center}}
\newcommand{\ii}{\ensuremath{\mathrm{i}}}
\def\lsim{\mathrel{\rlap{\lower4pt\hbox{\hskip1pt$\sim$}}
    \raise1pt\hbox{$<$}}}         
\def\gsim{\mathrel{\rlap{\lower4pt\hbox{\hskip1pt$\sim$}}
    \raise1pt\hbox{$>$}}}         

\newcommand{\ucharles}{Faculty of Mathematics and Physics, Charles University, V Hole\v{s}ovi\v{c}k\'ach 2, 18000 Prague 8, Czech Republic}
\newcommand{\uach}{Instituto de Ciencias F\'isicas y Matem\'aticas, Universidad Austral de Chile, Casilla 567, 5090000 Valdivia, Chile}
\newcommand{\infn}{Dipartimento di Matematica e Informatica, via Ospedale 72, 09123, Cagliari, Italy and INFN Sezione di Cagliari, Italy}
\newcommand{\borisaff}{Wohlergasse 6, 1100 Vienna, Austria}

\begin{document}

\title{The three ``layers'' of graphene monolayer \\
and their {analog} generalized uncertainty principles}
\author{A.~Iorio}
\email{alfredo.iorio@mff.cuni.cz}
\affiliation{\ucharles}
\author{B.~Iveti\'{c}}
\email{bivetic@yahoo.com}
\affiliation{\borisaff}
\author{S.~Mignemi}
\email{smignemi@unica.it}
\affiliation{\infn}
\author{P.~Pais}
\email{pais@ipnp.troja.mff.cuni.cz}
\affiliation{\uach}
\affiliation{\ucharles}


\date{\today}

\begin{abstract}
We show that graphene, in its simplest form and settings, is a practical table-top realization of the analog of exotic quantum gravity scenarios, which are speculated to lead to certain generalized Heisenberg algebras. In particular, we identify three different energy regimes (the ``layers'') where the physics is still of a pseudorelativistic (Dirac) type but more and more sensitive to the effects of the lattice. This plays here a role analog to that of a discrete space, where the Dirac quasiparticles live. This work improves and pushes further earlier results, where the physical meaning of the high energy momenta was clear, but the conjugate coordinates only had a purely abstract description. Here we find the physical meaning of the latter by identifying the mapping between the high-energy coordinates and low-energy ones, i.e., those measured in the lab. We then obtain two generalized Heisenberg algebras that were not noticed earlier. In these two cases, we have the striking result that the high-energy coordinates just coincide with the standard ones, measured in the lab. A third generalized Heisenberg algebra is obtained, and it is an improvement of the results obtained earlier in two respects: we now have an expression of the generalized coordinates in terms of the standard phase-space variables, and we obtain higher order terms. All mentioned results clearly open the doors to table-top experimental verifications of many generalized uncertainty principle-corrected predictions of the quantum gravity phenomenology.
\end{abstract}

\maketitle

\section{Introduction}
\label{intro}

It has been about a decade since Dirac materials \cite{wehling}, like graphene \cite{GeimNovoselov2005}, have been recognized as candidates for table-top analog realizations of a wide range of phenomena of high energy physics \cite{weylgraphene}. These include various conformal (Weyl) Killing horizons of the Bañados-Teitelboim-Zanelli (BTZ) black hole kind, of the Rindler kind, of the de Sitter kind, all related to some sort of Hawking-Unruh phenomenon \cite{i2}, as well as spatiotemporal torsion \cite{ip3,ip4}, the unconventioal supersymmetry (USUSY) \cite{AVZ,*APZ,*GPZ,*ADVZ2021} (see, e.g., \cite{DauriaZanelli2019} and also \cite{ip3}), and more \cite{GUPBTZ}. See \cite{Acquaviva:2022yiq} for a recent review.
	
In \cite{ip2,ip5} there is a first attempt to explore, in this context, the breakdown of the continuum structure of space at the smallest scales. The modification of the Heisenberg algebra for the conductivity electrons of graphene has been studied there, and in \cite{Jellal2021}. In contrast to the fundamental case, the deformation scale of the Heisenberg algebra here is the lattice spacing\footnote{One could also take as characteristic length, the ``elementary size'' of the topological defects naturally occurring in the plastic deformations of the membrane, such as dislocations, disclinations or grain boundaries \cite{ip3}. The orders of magnitude remain, of course, gigantic compared to Planck's.}, $\ell_{graphene} \sim \SI{e-10}{\metre}$, that, of course, is way larger than $\ell_{P}\sim \SI{e-35}{\metre}$, the Planck length.

This paper aims to pave the way for using experimental availability and manipulability of graphene to learn how the generalized dynamics leading to (or stemming from, depending on the point of view) these deformations are to be formulated. If this goal is reached, we shall have a real system at our disposal where we can study the effects of granularity of space on the dynamics of a Dirac quantum system. This last seems remarkable to us, as a crucial experimental window into the generalized uncertainty principles (GUPs) \cite{Maggiore:1993rv,Kempf:1993bq,Scardigli:1999jh,Buoninfante:2019fwr,Petruzziello:2020wkd,Bosso:2021koi} might be opened up in this arena. Some interesting results in this direction, where the role of graphene is addressed, but the discussion is more general, are in \cite{GUPBTZ}.

Let us remark here that an analog system cannot behave as the target system in all respects, but only in certain specific aspects, and it is important to clarify, case by case, what such aspects are. In other words, results obtained in an analog system cannot be \textit{universal}, as one expects, for instance, in relation to a \textit{fundamental} GUP. Indeed, GUPs refer to properties of the spacetime itself, henceforth, \textit{all} types of particles propagating in such ``GUP-compatible'' space should enjoy the novel properties. Of course, we cannot have that here.

Our only possible goal here is to reproduce the very same conditions that the analog of one very specific type of matter (Dirac massless fermions in $2+1$ dimensions) should be experiencing, when propagating in this ``GUP-compatible'' space. We cannot ask more to this simple system, but this would be already quite a precious result, in an area of research where all results are just theoretical or speculative.

To deal with this lack of universality, we have to duly take into account that particles, other than the Dirac quasi-particles of graphene, might not suitably behave as the analog of particles in presence of analog GUPs.

The typical scenario that we have in mind to reproduce here is the one for which the commutation relations are modified, by quantum gravity (QG) effects, to be (see, e.g., \cite{AliDasVagenas1,*AliDasVagenas2,*AliDasVagenas3,*AliDasVagenas4,*AliDasVagenas5} and references therein)
\be\label{ADV comm rel}
[x_i,p_j] = \ii \hbar \left( \delta_{i j} - A \left( |\vec{p}| \delta_{i j} + p_i p_j / |\vec{p}| \right) + A^2 \left( |\vec{p}|^2 \delta_{i j} + 3 p_i p_j  \right) \right) \;,
\ee
where $A = \tilde{A} \, \ell_P / \hbar$, with $\tilde{A}$ a phenomenological dimensionless parameter. As said, our scales here are much more within reach, as $\ell_{graphene} \sim 10^{25} \, \ell_P$, hence we have much more hope to be able to measure the effects of such modifications.

Some steps in this direction were taken in \cite{ip2}, where the commutation relations shown above were obtained, at $O(A)$ only, but for \textit{generalized} phase-space variables. Here we move further and look for the physical meaning of such generalized coordinates, which, in turn, gives meaning to those commutation relations. Most importantly, this will teach us in terms of which ``low energy'' (i.e., reachable and measurable at our scales) variables can realize such ``high energy'' granular dynamics. Indeed, such a split into ``high energy'', $(x_i,p_i)$, and ``low energy'',$(x_{0 i}, p_{0 i})$, phase-space variables is precisely what is behind the expression (\ref{ADV comm rel}), when
\be
p_i = p_{0 i} \left( 1 - A |\vec{p_0}| + 2 A^2 |\vec{p_0}|^2 \right) \,,
\ee
and $x_i = x_{0 i}$, with $[x_{0 i}, p_{0 j}] = \ii \hbar \delta_{i j}$, see \cite{AliDasVagenas1,*AliDasVagenas2,*AliDasVagenas3,*AliDasVagenas4,*AliDasVagenas5} and references therein.

In our case, we actually have more than one ``high energy layer''. In fact, in principle, we might have \textit{infinite} such ``layers'', related to the infinite ladder of contributions from further and further atoms of the lattice (near neighbors, next-to-near neighbors, next-to-next-to-near neighbors, etc), see \cite{ip2} and Section \ref{SnyderGraphene}. Such ``layers'' have to be taken into account as the energy and the effects of granularity grow away from the ``low energy layer'', located near the Fermi points. Nonetheless, as explained in \cite{ip2} and later here, in order to retain at least some of the Dirac Hamiltonian structure, we are forced to stop at \textit{two} such ``high energy layers'', making up, altogether, the ``three layers'' we refer to in the title of this paper.

As we shall see, starting from the phase-space variables of the low energy set-up, that are somehow fixed, we have various natural choices for the two other ``layers''. The analysis that we carry on from there, on the one hand, leads to \textit{three} kinds of GUPs: one for the first ``high energy layer'', that went unnoticed in earlier research; two for the second ``high energy layer'', one of which also went unnoticed while the other is an improvement of earlier findings. On the other hand, we solve the important problem of how to relate the high energy coordinates with the low energy phase-space variables. This latter point is of paramount importance to set-up experiments and identify the measurable effects of such deformations on the Dirac Hamiltonian descriptions. We shall not do that here, but should give a general recipe in the concluding remarks.

One important result here is that, in two important cases with GUPs, such high energy coordinates simply coincide with the low-energy coordinates. In other cases, such coordinates have complicated expressions in terms of the standard, measurable phase space variables.

As for the momenta, their high-energy expressions is given once and for all in terms of the low-energy, measurable momenta, by taking into account the effect of the dispersion relations at higher and higher order. We may say that it is the physics that dictates these expressions.

Once these expressions are under control, when we measure this or that observable, even though we only have access to low-energy momenta, we simply need to recognize the generalized momenta (and, when necessary, the generalized coordinates) in the given expression. This allows us to experimentally see the effects of the GUPs on the physics of the Dirac quasi-particles, which appears to us quite remarkable.

The paper is organized as follows. In \Cref{SnyderGraphene} we revisit graphene dispersion relations, so we shall build our construction based on the experimental facts reported in this phenomenological Section. Such constructions are then presented in \Cref{Section VarSuperHyper}, where the three ``layers'' are clearly explained. In \Cref{Sect_from_first_to_second_layer} we find the transformations leading from the first ``layer'' to the second ``layer'', assuming that the standard variables are always canonical, while in \Cref{Sect_from_second_to_third_layer} we move from the second ``layer'' to the third ``layer'', always keeping contact with the first ``layer'' that makes the expressions we obtain for the coordinates something with a definite physical meaning. The last Section is devoted to the conclusions, and we also offer three appendices with more details of the more involved calculations.

\section{Graphene dispersion relations revisited}
\label{SnyderGraphene}

Dirac materials \cite{wehling}, like graphene \cite{PacoReview2009}, have two-dimensional hexagonal lattice structures made of two interpenetrating triangular sublattices, usually indicated as $L_A$ and $L_B$, with associated fermionic annihilation and creation operators, $(a_{\vec{k}}, a^*_{\vec{k}})$ and $(b_{\vec{k}}, b^*_{\vec{k}})$, respectively. Their tight binding complete field Hamiltonian is
\be
{\cal H}_{\vec{k}} = \sum_{m \in \textrm{diag}} \eta_m   {\cal F}_m ({\vec{k}}) (a^*_{\vec{k}} a_{\vec{k}} + b^*_{\vec{k}} b_{\vec{k}})
+ \left( \sum_{m \in \textrm{off}} \eta_m  {\cal F}^*_m ({\vec{k}}) a^*_{\vec{k}} b_{\vec{k}} + h.c. \right)   \label{fieldHthird}
\ee
where all vectors are two dimensional, ${\vec{k}} = (k_x, k_y)$, $\eta_m$ are the hopping parameters, $m = 1, 2, ...$ and the first sum is for hopping from an atom of a given sublattice to atoms sitting in the same sublattice, while the second sum is for hopping to atoms belonging to the other sublattice. Each function
\be
{\cal F}_m ({\vec{k}}) \equiv \sum_{i=1}^{n_m} e^{\ii {\vec{k}} \cdot {\vec{s}^{(m)}}_i} \,,
\ee
encodes the relevant information on the geometry of the lattice relative to the $m^{\rm th}$-near neighbors. The details on the derivation of (\ref{fieldHthird}) and on the ${\cal F}_m$s are in \cite{ip2}, where it is explained that the $n_m$ vectors $\vec{s}^{(m)}_i$ connect any given atom to its $m^{\rm th}$-near neighbors.

As is well known, the function for the nearest neighbors, $m=1$, can be written as \cite{PacoReview2009}
\begin{equation}\label{firstfunction}
  {\cal F}_1 ({\vec{k}}) = \sum_{i= 1}^{3} e^{\ii {\vec{k}} \cdot {\vec{s}^{(1)}}_i} =
  e^{- \ii \ell k_y} \left[ 1 + 2 e^{\ii \frac{3}{2} \ell k_y} \cos\left(\frac{\sqrt{3}}{2} \ell k_x\right) \right] \,.
\end{equation}
Here and in what follows we shall call the lattice spacing $\ell$.

Among the results of the analysis of \cite{ip2} is that, due to the specific geometric structure of the material, the function ${\cal F}_2 (\vec{k})$, encoding the information on the next-to-near neighbors, can be expressed fully in terms of ${\cal F}_1 (\vec{k})$:
\be
{\cal F}_2 = |{\cal F}_1|^2 - 3 \,,
\ee
while similar relations do not hold for higher values of $m$. This has an impact on the dispersion relations, that for us is important. In fact the latter descends from the secular equation, ${\rm det}   \left( {\cal H}_{\vec{k}} - E  {\cal S}_{\vec{k}} \right) = 0$, where ${\cal S}_{\vec{k}}$ is the overlapping matrix.

Since there is no simple way to express ${\cal F}_m ({\cal F}_1)$, with the exception of $m=2$, and given that the physics is well described by near and next-to-near neighbor interactions, we stop at $m=2$ to get, as the two solutions of the eigenvalue equation, expressions fully given in terms of ${\cal F}_1$ alone
\begin{equation}\label{fullDispOrd2}
E_\pm = \frac{\epsilon_0 - \epsilon_1 |{\cal F}_1|^2 \pm \sqrt{(\epsilon_0 - \epsilon_1 |{\cal F}_1|^2)^2
- ( 1 - \varsigma_1^2 |{\cal F}_1|^2) [(\epsilon_0 + \eta_2 |{\cal F}_1|^2)^2 - \eta^2_1 |{\cal F}_1|^2 ]}}{ 1 - \varsigma_1^2 |{\cal F}_1|^2} \,,
\end{equation}
where $\epsilon_0 \equiv \eta_0 - 3 \eta_2$, $\epsilon_1 \equiv \varsigma_1 \eta_1 - \eta_2$, and $\varsigma_1$ is the first of the overlapping parameters, $\varsigma_m$, of ${\cal S}_{\vec{k}}$.

What has been said up to here holds for any member of the family of two-dimensional Dirac materials \cite{wehling}. For definiteness, let us consider now the ideal case of an infinite and non-deformed sheet of graphene, and let us give here the values of the first few parameters of importance (see, e.g., \cite{ferreira} and references therein):
\bea
\eta_0 \simeq - \SI{0.36}{\electronvolt} \,, & \, \varsigma_1 \simeq 0.106 \,, & \epsilon_0 \simeq -\SI{0.72}{\electronvolt} \,, \nonumber \\
\eta_1 \simeq - \SI{2.8}{\electronvolt} \,,  & \, \varsigma_2 \simeq 0.001 \,, & \epsilon_1 \simeq - \SI{0.43}{\electronvolt} \,, \label{parameters} \\
\eta_2 \simeq \SI{0.12}{\electronvolt} \,,   &  & \nonumber
\eea
from which we see that stopping at $\eta_2$ and $\varsigma_1$ is a very good approximation.

Let us now set the zero of the energy at the Dirac point, $E_\pm|_{k_D} = 0$, i.e., $E_\pm \to E_\pm - \epsilon_0$, and let us expand (\ref{fullDispOrd2}),
\bea
  E_\pm &\simeq& \left(1+0.012 |{\cal F}_1|^2\right) \left[ \pm \eta_1 |{\cal F}_1| \left( 1 - 0.028 + 0.0047 |{\cal F}_1|^2\right) - \epsilon_1 |{\cal F}_1|^2 \right] \nonumber \\
  &\simeq& \eta_1 \left( \pm 0.97 |{\cal F}_1| - 0.15 |{\cal F}_1|^2 \pm 0.017 |{\cal F}_1|^3\right) \nonumber \\
  &\simeq& V_F \left( \pm P_0 - A  P_0^2 \pm B^2 P_0^3 \right) \label{DispRelModified}
\eea
where
\be \label{VF}
V_F \simeq 0.97 \, \frac{\eta_1 \ell}{\hbar} \,,
\ee
is a velocity parameter (recall that the Fermi velocity is $v_F = 3/2 \, \eta_1 \, \ell / \hbar$, so it is about $50$ percent bigger \cite{PacoReview2009})
and
\be \label{A}
A \simeq 0.15 \, \frac{\ell}{\hbar}
\ee
and $B \simeq 0.13 \, \ell / \hbar$. The most important definition, though, is that of the two-dimensional vector, with the dimension of momentum,
\be\label{supermomenta}
{\vec{P}_0} \equiv - \frac{\hbar}{\ell} \, \left( {\rm Re} {\cal F}_1, {\rm Im} {\cal F}_1 \right) \,,
\ee
whose length we indicate with $|\vec{P_0}|$. Expanding $|\vec{P_0}|$, one has \cite{PacoReview2009}
\be \label{P0(p)polarcoord}
|\vec{P_0}| \sim \alpha \, |\vec{p}| + \beta \, \ell \, |\vec{p}|^2 + \gamma \, \ell^2 \, |\vec{p}|^3 \,,
\ee
with $\vec{p}$ a small momentum around the Dirac points (see \cite{PacoReview2009} and below) and  $\alpha, \beta, \gamma$ real coefficients. Inserting this into (\ref{DispRelModified}), and {\it dropping $O(\ell^3)$ terms,}\footnote{\label{footnoteEllP} One should recall that there is always an overall factor $O(\ell)$, given by the Fermi velocity in (\ref{VF}), or later in $v_F \sim 3/2 \, \eta_1/\hbar  \, \ell$, so that $O(p^m)$ and $O(\ell^m)$, in the various expansions of the energy and of the Hamiltonian, go actually together.} one sees that the physics at $O(\ell^2)$ is well described if one simply keeps the first two terms in the expansion, i.e.,
\be
E_\pm  =   V_F \left( \pm |\vec{P_0}| - A \; |\vec{P_0}|^2 \right) \label{DispRelLargeP} \,.
\ee
Notice that in the following we shall use a slightly different definition of $\vec{P_0}$, that allows one to expand it as $\vec{P_0} \sim \vec{p} + \cdots $, rather than $\vec{P_0} \sim 3/2 \vec{p} + \cdots $.
That is $\vec{{\cal P}_0} \equiv 2/3 \vec{P_0}$, but to simplify notation with an abuse we shall simply write $\vec{{\cal P}_0} \to \vec{P_0}$.

Some comments are now in order. This is a doubly folded situation. On the one hand, one can decide at what order of neighbors, $m$, one wants to operate: nearest neighbor, next to nearest, etc. The typical choice in the literature is $m=1$. Here, and in \cite{ip2}, we moved to $m=2$. What counts, though, is at what order in powers of $\ell$ (or $p$, see footnote \ref{footnoteEllP}) we operate, and this  needs always to be specified, no matter the choice of $m$. Henceforth, when an expression is given in terms of $P_0$, one needs to expand the latter as $P_0(p)$, at the given order in $\ell$, because what we are really looking for is the $O(\ell^2)$ effects, that are as well in $O(A)$, and so on.

The dispersion relation (\ref{DispRelLargeP}) can be obtained from the secular equation, ${\rm det} (H - E\mathbb{1}) = 0$, for the following Hamiltonian, written in terms of the supermomenta $P_{0}$,
\begin{equation}\label{newDiracHwithP0}
  H \equiv \sum_{\vec{k}} {\cal H}_{\vec{k}} =
  V_F  \sum_{\vec{k}} \psi^\dag_{\vec{k}} \left(\not\!P_0 - A \; \not\!P_0 \not\!P_0 \right) \psi_{\vec{k}} \,,
\end{equation}
where $\slashed{P}_0 \equiv \vec{\sigma} \cdot {\vec{P}_0}$ and our convention is \footnote{There are many conventions related to different choices for the pairs of inequivalent Dirac points, and different arrangements of the $a$ and $b$ operators to form the spinor $\psi$. See Appendix B of \cite{ip3} for more details.} $\psi_{k}^{\dagger}=(b_{k}^{*}, a_{k}^{*})$, where $a_{k}$ and $b_{k}$ are the annihilation operators for the $L_A$ and $L_B$ sublattices, see \cite{Iorio2013}.

Noticeably, this is the same Hamiltonian that is obtained in the phenomenology of QG, when generalizing the Dirac Hamiltonian, to accommodate a GUP with a minimal fundamental length \cite{AliDasVagenas1,*AliDasVagenas2,*AliDasVagenas3,*AliDasVagenas4,*AliDasVagenas5}. Here such a fundamental length is clearly given by the carbon to carbon distance $\ell$.

In those papers \cite{AliDasVagenas1,*AliDasVagenas2,*AliDasVagenas3,*AliDasVagenas4,*AliDasVagenas5}, the authors introduce \textit{two} different ``layers'' of the description of the given physical system. Namely, ``high energy'' and ``low energy'' phase-space variables, that, adjusting our notation to theirs, we may say call $(X,P)$ and $(X_0, P_0)$, respectively. Here (and there) the definition of the ``high energy momenta'' descends directly from (\ref{newDiracHwithP0}), which in turn is associated to the dispersion relations (\ref{DispRelLargeP})
\begin{equation}\label{hypermomenta}
{\vec{P}} \equiv {\vec{P}_0} (1 - A |\vec{P_0}| ) \,,
\end{equation}
and, if one then uses the customary Dirac prescription, $|\vec{P_0}| \to \vec{\sigma} \cdot \vec{P_{0}}$, this clearly gives
\be \label{H(P)}
H = V_F \sum_{\vec{k}} \psi^\dag_{\vec{k}} \slashed{P} \psi_{\vec{k}} \,,
\ee
as the effective Hamiltonian at this higher energy ``layer''.

With a slightly different notation, most of what was just said was seen already in \cite{ip2}. There were some crucial problems that were left open, though, and it is our intention to solve them here. Among these problems, the most important is the physical meaning of the \textit{abstract} coordinates $\vec{X}_0$ and $\vec{X}$. In fact, while in \cite{AliDasVagenas1,*AliDasVagenas2,*AliDasVagenas3,*AliDasVagenas4,*AliDasVagenas5}, and in the related literature, such coordinates are nothing other than the ones we measure at our energy scales (in other words, $\vec{X} = \vec{X}_0$, and no other scale is present), in our case the situation is actually richer. As we shall explain in the following section, this richer scenario is here because we have a third scale, and  such a scale is actually the most important of all.

Before turning our attention to that, let us stress that it is only when we know the physical meaning of the coordinates involved, that we can use the system to experimentally reproduce the exotic GUP scenarios of the theoretical literature, such as, e.g., GUP-induced corrections to the Klein tunneling (see \cite{AliDasVagenas1,*AliDasVagenas2,*AliDasVagenas3,*AliDasVagenas4,*AliDasVagenas5} for the theoretical case, and \cite{KleinGraphene} for the Klein tunneling on graphene), etc.
\begin{figure}
{\includegraphics[width = 5in]{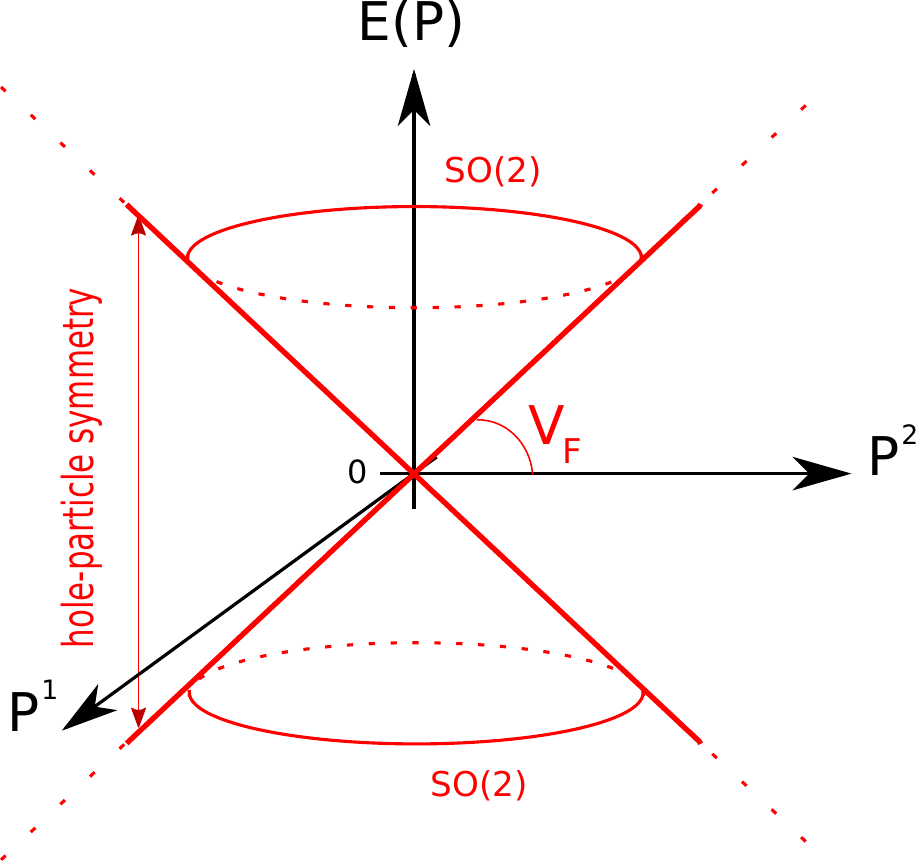}}
\caption{This plot shows the dispersion relations as seen from the fully symmetric ``layer'', the hypervariables $(X,P)$, that is $E (P) = \pm V_F |\vec{P}|$. Here the slope is $V_F$ and the Dirac cones are undeformed and perfectly particle-hole symmetric. If we keep going beyond the Dirac point, to fill the whole Brillouin zone, these variables are just extending the Dirac cone structure all over, as indicated with the dashed straight lines. The actual dispersion relations are both below (at the start the real slope is constant and $v_F < V_F$) and above (then the slope becomes smaller than $V_F$ and non-constant), see, e.g., \cite{PacoReview2009}, so that this is a sort of ``average'' behavior, as seen from the highest energy ``layer'' of the system. See also the next two Figures, especially Fig. \ref{fig:a3}. In this plot $\hbar = 1 = \ell$ and the scale is such that $V_F = 1$.}
\label{fig:a1}
\end{figure}

\section{Variables, Supervariables and Hypervariables}
\label{Section VarSuperHyper}

\begin{figure}
{\includegraphics[width = 5in]{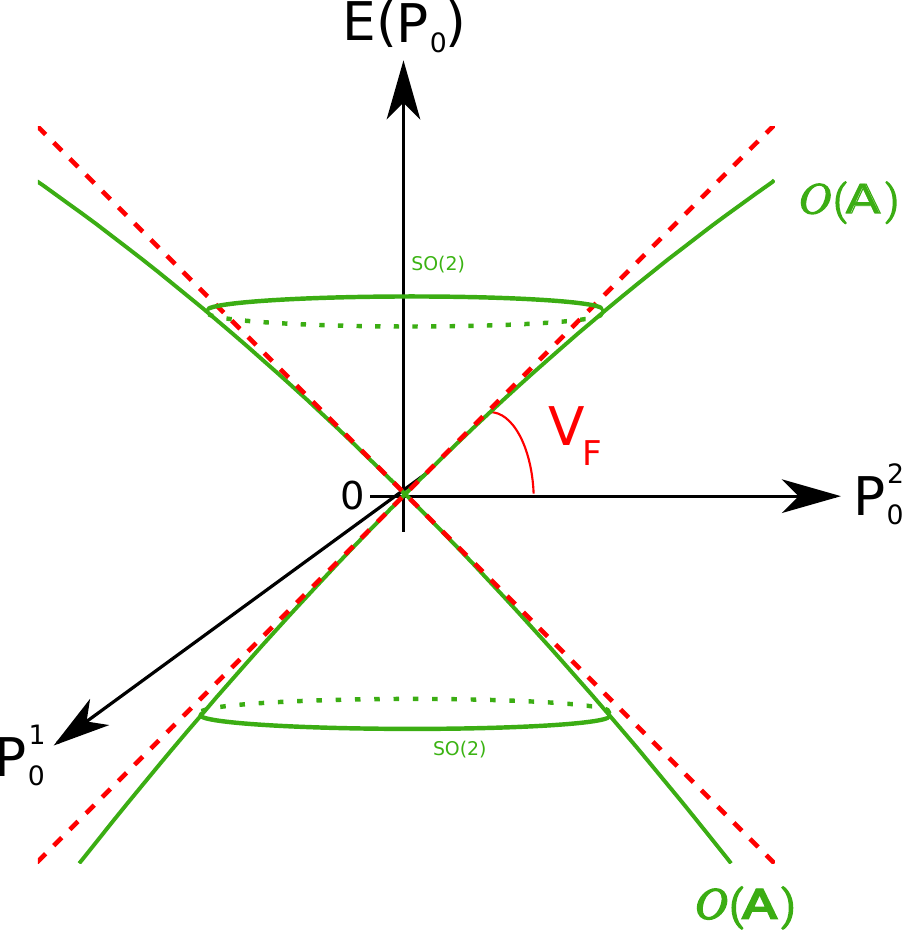}}
\caption{This plot shows the dispersion relations as seen from the middle-way symmetric ``layer'', the supervariables $(X_0,P_0)$, that is $E (P_0) = V_F \left( \pm |\vec{P_0}| - A \; |\vec{P_0}|^2 \right)$, given in Eq. (\ref{DispRelLargeP}). Here the initial slope is still $V_F$, but the Dirac cones are deformed, hence the slope is actually non-constant and the particle-hole symmetry is gone. This is not seen in the immediate neighbor of the Dirac points, where the Dirac cones are still preserved till the contribution of the second order terms, $|P_0|^2$, becomes important. When we keep going beyond the linear approximation, the non-constant nature of the slope becomes more and more visible, as shown. The SO(2) symmetry, though, is preserved all along the Brillouin zone, that of course is an approximation of the real behavior, but it is a better approximation compared to the one given by the hypervariables ($X,P$), see Fig. \ref{fig:a1}, and not as good as the approximation obtained using the variables ($x,p$), see Fig. \ref{fig:a3}. Here too $\hbar = 1 = \ell$, so that $A = 0.15$, and $V_F = 1$.}
\label{fig:a2}
\end{figure}

\begin{figure}
{\includegraphics[width = 5in]{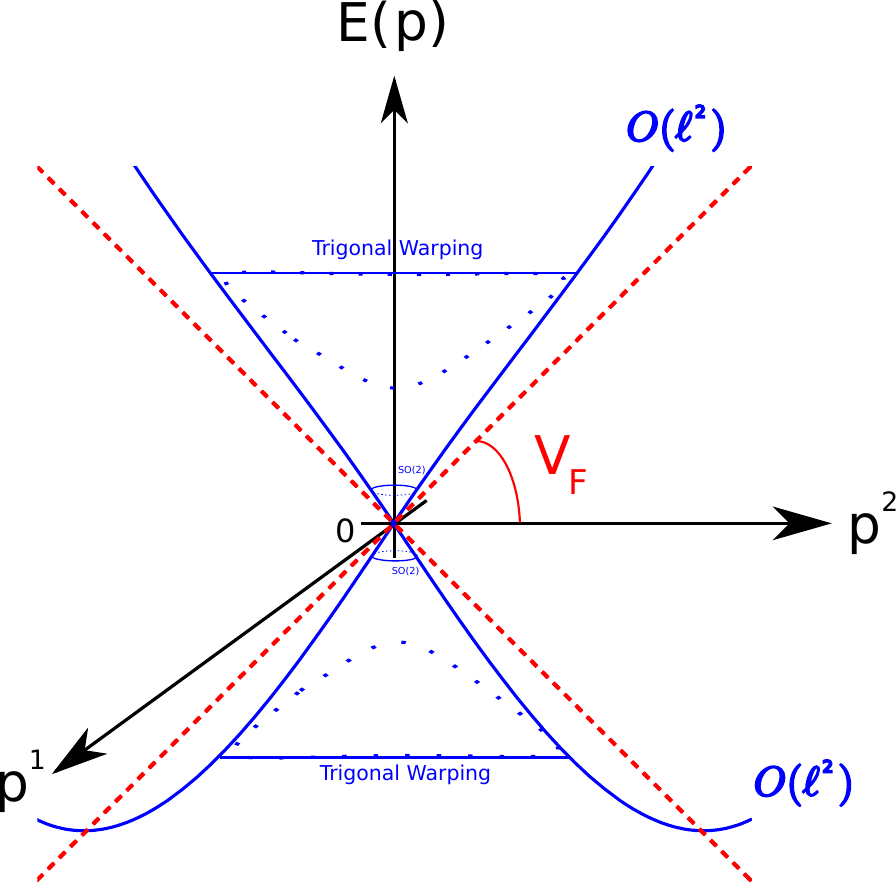}}
\caption{This plot shows the dispersion relations as seen from the least symmetric ``layer'', the variables $(x,p)$, that is those given in Eq. (\ref{DispRelsmallp}), obtained by substituting $P_0 (p)$. The initial slope is now $v_F \simeq 1.5 V_F$ and the Dirac cones are deformed not only as in the suparvariables layer but they also become conoids with a triangular $\mathbb{Z}_3$ symmetry. Thus, in this case, the SO(2) symmetry is also gone, along with the particle-hole symmetry. Again, this is not seen in the immediate neighborhood of the Dirac points, where the Dirac cones are still preserved, SO(2) symmetric and particle-hole symmetric, till the contribution of the $O(\ell^2)$ terms becomes important. These dispersion relations are much closer to the real dispersion relations of graphene quasiparticles than the other two approximate expressions given in Fig. \ref{fig:a1} and \ref{fig:a2}, see, e.g., \cite{PacoReview2009}. Since we can still recognize a Dirac structure in this layer, the complicated and non-symmetric behavior depicted here is actually an invaluable richness to reproduce QG effects on the Dirac theory. As before, $\hbar = 1 = \ell$, $A = 0.15$, and $V_F=1$, and here we also have $v_F = 1.5$.}
\label{fig:a3}
\end{figure}

As announced earlier, the main feature of the system in our hands is that here we have \textit{three} ``layers'', rather than the two layers of the scenario descending from QG phenomenology \cite{AliDasVagenas1,*AliDasVagenas2,*AliDasVagenas3,*AliDasVagenas4,*AliDasVagenas5}. Here we have
\begin{equation}\label{layers}
  (x,p) \,, \quad (X_0, P_0) \,, \quad (X, P) \,,
\end{equation}
in order of increasing energy necessary to use the given description. These three ``worlds'' manifest themselves through the dispersion relations, as indicated in detail in \Cref{fig:a1,fig:a2,fig:a3}. The \textit{variables} $(x,p)$ are the actual phase-space variables of the lowest energy Dirac quasiparticles of graphene, which originate from the $O(\ell)$ dispersion relations. These are the variables used in all papers relying on the Dirac description of the conductivity electrons of graphene, from the early days of \cite{semenoff1984}, untill the recent \cite{Jellal2021}, and including the papers on analog high energy physics, see, e.g. \cite{weylgraphene,i2,ip3,ip4,DauriaZanelli2019}.

Before going on, we should clarify that even if we started with a \textit{second-quantization picture} (by defining $\psi$ in terms of creation and annihilation operators for $A$ and $B$ sublattices), from now on we use the \textit{one-particle} or \textit{first-quantization picture}. It is for that reason that the operators are the variables (also supervariables or hypervariables) but not $\psi$, which should be interpreted as a wave function of the $\pi$ electron. These variables have a clear physical meaning. They are the (quantum mechanical) coordinates and momenta of the low energy conductivity electrons of graphene, whose dynamics occurs (quantum mechanically) on the continuum two-dimensional \textit{membrane}, which is how electrons of such energy/wavelength see the (hexagonal) lattice.

There are then the two further sets of phase-space variables we have seen earlier. We shall now call them \textit{supervariables}, $(X_0, P_0)$, and \textit{hypervariables}, $(X, P)$, as both sets refer to functions that appear in the dispersion relations that contain \textit{all orders} in $\ell$
\begin{equation}\label{disprel_1}
  E_\pm  \sim \eta_1 (\pm |{\cal F}_1| - \tilde{A} |{\cal F}_1|^2) \;,
\end{equation}
where $\tilde{A} \simeq 0.15$ is the dimensionless phenomenological parameter related to $A$ in (\ref{A}). While it is literally true that the function ${\cal F}_1$, in (\ref{firstfunction}), contains all orders in $\ell$, as recalled in the previous section (see also \cite{ip2}), one needs to decide at which order one would like to proceed, say $O(\ell^m)$, and then all quantities need be taken at that order $m$. Furthermore, the simple expression (\ref{disprel_1}) [see also (\ref{DispRelModified})] is already an approximation that holds when higher order contributions are not considered.

Thus, what we are saying here is that (writing things in a dimensionless fashion, i.e., not including a necessary factor of $1/\ell$)
\begin{equation}\label{3layersmomenta}
|{\cal F}_1| \sim p + O(\ell^2) \,, \quad |{\cal F}_1| \sim P_0 \,, \quad  |{\cal F}_1| - \tilde{A} |{\cal F}_1|^2 \sim P \,,
\end{equation}
where the \textit{supermomenta}, $\vec{P}_0$, are given in (\ref{supermomenta}) and the \textit{hypermomenta}, $\vec{P}$, are given in (\ref{hypermomenta}).

These relations are key in our analysis and are dictated by the phenomenology of the system. Their relation to the measurable momenta $\vec{p}$ is known, at any order. What remains to be found is the relation of supercoordinates and hypercoordinates to the variables, $(x,p)$.

For the sake of clarity and of the calculations, which we shall carry on in a moment, let us consider explicitly the near-neighbors contributions, $m=1$. The Hamiltonian reads (we use $\hbar = 1$ from here on)
\begin{equation}\label{Hamiltonian_near_neighbors_a_b}
H = \eta_1 \sum_{\vec{k}} \left( {\cal F}_1 ({\vec{k}}) a^{*}_{\vec{k}} b_{\vec{k}} + H.c. \right)
\end{equation}
where ${\cal F}_1$ is the complex function in (\ref{firstfunction})
\[
  {\cal F}_1 ({\vec{k}}) = \sum_{i= 1}^{3} e^{\ii {\vec{k}} \cdot {\vec{s}^{(1)}}_i} =
  e^{- \ii \ell k_y} \left[ 1 + 2 e^{\ii \frac{3}{2} \ell k_y} \cos\left(\frac{\sqrt{3}}{2} \ell k_x\right) \right] \,.
\]
To see how the $(2+1)$-Dirac Hamiltonian arises, we need to expand ${\cal F}_1$ around one of the two inequivalent Fermi points, $K^{D}_{\pm}=(\pm\frac{4\pi}{3\sqrt{3}\ell},0)$. So, if $k_{i}=K^{D}_{+i} + p_{i}$, the Hamiltonian (\ref{Hamiltonian_near_neighbors_a_b}), up to $O(\ell^{3})$, is \cite{ip2}
\begin{eqnarray}\label{Hamiltonian_near_neighbors}
H & = & v_F \sum_{\vec{p}} \psi^\dag_{\vec{p}} \Bigg{[} \sigma_1 \left( p_1 - \frac{\ell}{4} (p^2_1 - p^2_2) - \frac{\ell^2}{8} p_1 (p^2_1 + p^2_2) \right) \nonumber \\
  & + & \sigma_2 \left( p_2 + \frac{\ell}{2} p_1 p_2 - \frac{\ell^2}{8} p_2 (p^2_1 + p^2_2) \right) \Bigg{]} \psi_{\vec{p}} \;,
\end{eqnarray}
where the standard Dirac Hamiltonian is seen when only linear terms in $p$ are considered, $H = v_F \sum_{\vec{p}} \psi^\dag_{\vec{p}} \, \vec{\sigma} \cdot \vec{p} \, \psi_{\vec{p}}$. Notice that here $v_F = 3/2 V_F$ and $\sigma_{1}$, $\sigma_{2}$ are Pauli matrices. This is what defines the relations $P_0(p)$ of \Cref{SupermomentumCartesian}, below [see also the same relation (\ref{P0(p)polarcoord}), written in polar coordinates  in momentum space, $p_1 = |\vec{p}|\, \cos \theta$ and $p_2 = |\vec{p}|\, \sin \theta$].

Now we consider the next-to-near neighbors contributions, $m=2$, and we do that by considering the Hamiltonian (\ref{newDiracHwithP0}) that, at that given order, explicitly reads
\begin{eqnarray}\label{Hamiltonian_total_Op3}
H & = & v_F \sum_{\vec{p}} \psi^\dag_{\vec{p}} \Bigg{[} \sigma_1 \left( p_1 - \frac{\ell}{4} (p^2_1 - p^2_2) - \frac{\ell^2}{8} p_1 (p^2_1 + p^2_2) \right) \nonumber \\
  & + & \sigma_2 \left( p_2 + \frac{\ell}{2} p_1 p_2 - \frac{\ell^2}{8} p_2 (p^2_1 + p^2_2) \right) \nonumber \\
  & - & \frac{3}{2} A \left( (p^2_1 + p^2_2) - \frac{\ell}{2} p_1^3 + \frac{3 \ell}{2} p_1 p^2_2 \right) \Bigg{]}  \psi_{\vec{p}} \;.
\end{eqnarray}

The dispersion relations for the latter Hamiltonian are
\begin{equation}\label{DispRelsmallp}
  E_\pm   =  v_F \left( \pm |\vec{p}| \mp  \frac{\ell}{4} |\vec{p}|^2 \cos 3\theta \mp  \frac{\ell^2}{64}  |\vec{p}|^3 ( 7 + \cos 6\theta )
  -  \frac{3}{2} A\, |\vec{p}|^2 + \frac{3 }{4} A\,\ell\, |\vec{p}|^3
  \cos 3\theta \right) \;,
\end{equation}
where, to simplify the expression, we only used polar coordinates in momentum space.
Therefore, as seen in the general discussion of the last section, and more explicitly here, the interpretation of $P^i_0$ and $P^i$ as generalized momenta is suggested by the dispersion relations and by the form of the associated Hamiltonian, see the three figures and their captions. What is left to do is to find the meaning of the conjugate coordinates, $X^i_0$ and $X^i$, respectively.

In other words, we do know supermomenta and hypermomenta in terms of the measurable momenta
\begin{equation}\label{momenta}
P_{0}^{i}(p) \,\,\,\, \mbox{and} \,\,\,\, P^i (P_{0}(p)) \,,
\end{equation}
and we are looking for supercoordinates and hypercoordinates in terms of measurable phase-space variables $(x,p)$
\begin{equation}\label{coordinates}
X_{0}^{i}(x,p) \,\,\,\, \mbox{and} \,\,\,\, X^i (X_0 (x,p), P_{0}(p)) \,.
\end{equation}
Given the results of \cite{ip2}, we expect that this analysis will open the doors to noncanonical coordinate-momentum commutations, hence to GUPs. The crucial novelty here is that we shall search for the conditions that the generalized coordinates in (\ref{coordinates}) have to fulfill to have natural GUPs at the various layers. This way we shall have, on the one hand, a larger variety of cases, which earlier were not seen. On the other hand, we shall be able to say, case by case, what the generalized coordinates are in terms of measurable $(x,p)$, hence we shall be able to say what needs to be done, in practice, in order for the system to actually realize a dynamics in the presence of GUPs. In doing this analysis, we shall not explore the possibility of noncommuting coordinates.

\section{From first to second layer}
\label{Sect_from_first_to_second_layer}

In the first layer, $(x,p)$, we can take as an experimental fact that the phase-space variables are canonical (here $i,j=1,2$)
\begin{equation}\label{commutators_x_p}
[x^{i},p^{j}]= \ii\delta^{ij}  , \ \ \ [x^{i},x^{j}] = 0 = [p^{i},p^{j}] \;.
\end{equation}
Indeed, we deal here with pristine graphene, with no strain, hence no pseudomagnetic field and no external magnetic field, two instances that could produce noncommuting coordinates, see, e.g., \cite{ip5,Jackiw}.

The supermomentum $\vec{P_0} \equiv (P_0^1, P_0^2)$, at order\footnote{See the previous footnote.} $O(\ell^2)$ can be read off from the Hamiltonian (\ref{Hamiltonian_near_neighbors})
\begin{align}\label{SupermomentumCartesian}
P^1_0 & =  p^1 + \frac{\ell}{4} \left((p^2)^2 - (p^1)^2\right) -  \frac{\ell^2}{8} p^1 \left((p^1)^2 + (p^2)^2\right) \,, \nonumber \\
P^2_0 & =  p^2 + \frac{\ell}{2} p^1 p^2 -  \frac{\ell^2}{8} p^2 \left((p^1)^2 + (p^2)^2\right)  \,.
\end{align}
Although we shall always work with these expressions, let us incidentally present the following expression
\begin{align}\label{P_0 trigonal}
P^1_0 & =  p^1 - \frac{\ell}{4} |\vec{p}|^2 \cos 2 \theta -  \frac{\ell^2}{8} |\vec{p}| \cos \theta \;, \nonumber \\
P^2_0 & =  p^2 + \frac{\ell}{4} |\vec{p}|^2 \sin 2 \theta -  \frac{\ell^2}{8} |\vec{p}| \sin \theta \;,
\end{align}
where $\tan \theta = p^2/p^1$. This $\theta$ dependence is what, at higher energies, distorts the Dirac cone into a triangularly shaped conoid (see Fig. \ref{fig:a3}), a phenomenon called \textit{trigonal warping} \cite{PacoReview2009}.

From (\ref{SupermomentumCartesian}), with (\ref{commutators_x_p}), we have that
\begin{equation}\label{spurious commutations}
  [x^i , P_0^j(\vec{p})] = \ii F^{i j} (\vec{p}) \,,
\end{equation}
where
\begin{equation}\label{Fij(p)}
  F^{i j} (\vec{p}) = \delta^{i j}
  + \frac{\ell}{2}   \left(
                  \begin{array}{cc}
                  - p^1 & p^2 \\
                    p^2 & p^1 \\
                  \end{array}
                \right)
  - \frac{\ell^2}{8}  \left(
                  \begin{array}{cc}
                    3(p^1)^2 + (p^2)^2  & 2 p^1 p^2 \\
                    2 p^1 p^2 &  (p^1)^2 + 3(p^2)^2  \\
                  \end{array}
                \right) \,.
\end{equation}
Notice that $F^{ij}(\vec{p}) = F^{ji}(\vec{p})$, and that
\begin{equation}\label{traceF(p)}
    F^{i}_i (\vec{p}) = 2 - \frac{\ell^2}{2} |\vec{p}|^2 \,.
\end{equation}
We can then invert (\ref{SupermomentumCartesian})
\begin{eqnarray}
p^1 & = & P^1_0 - \frac{\ell}{4} \left((P^2_0)^2 - (P^1_0)^2\right) +  \frac{\ell^2}{4} P^1_0 \left((P^1_0)^2 + (P^2_0)^2\right) \,, \nonumber \\
p^2 & = & P^2_0 - \frac{\ell}{2} P^1_0 P^2_0 +  \frac{\ell^2}{4} P^2_0 \left((P^1_0)^2 + (P^2_0)^2\right) \,, \label{p(P_0)}
\end{eqnarray}
to obtain
\begin{equation}\label{Fij(P0)}
  F^{i j} (\vec{P_0}) = \delta^{i j}
  + \frac{1}{2} \, \ell \, \left(
                  \begin{array}{cc}
                  - P^1_0 & P^2_0 \\
                    P^2_0 & P^1_0 \\
                  \end{array}
                \right)
  - \frac{1}{2} \, \ell^2 \left(
                  \begin{array}{cc}
                    (P^1_0)^2  & P^1_0 P^2_0 \\
                    P^1_0 P^2_0 &  (P^2_0)^2   \\
                  \end{array}
                \right) \,.
\end{equation}
Similarly to the quantity $F^{ij}(\vec{p})$, we have $F^{ij}(\vec{P_{0}})  = F^{ji}(\vec{P_{0}})$, and
\begin{equation}\label{traceF(P0)}
    F^{i}_i (\vec{P_0}) = 2 - \frac{\ell^2}{2} |\vec{P_0}|^2 \,.
\end{equation}

We then have various choices for the supervariables, $(X_0,P_0)$, which require, in general, that we promote
\begin{equation}\label{X_0 general}
x^i \to X^i_0 (x, p) \,.
\end{equation}
Of course, once we are at a given layer, $(X_0,P_0)$ here, we must reexpress everything in terms of the appropriate phase-space variables. Nonetheless, the expression (\ref{X_0 general}) is very important, because it will establish a link of the supercoordinates, $X^i_0$, with the measurable variables $(x,p)$.

The two most natural choices are
\begin{equation}\label{Super Noncommutative}
  [X_0^i , P_0^j] = \ii F^{i j} (\vec{P_0}) , \ \ \ [X_0^i , X_0^j] = \ii G^{i j} , \ \ \ [P_0^i , P_0^j] = 0  \,.
\end{equation}
and
\begin{equation}\label{Super Canonical}
  [X_0^i , P_0^j] = \ii \delta^{i j}  , \ \ \ [X_0^i , X_0^j] = 0 = [P_0^i , P_0^j]   \,,
\end{equation}
which we shall call the ``GUP choice'' and the ``canonical choice,'' respectively. In fact, since $P^i_0(p)$ and (\ref{commutators_x_p}) hold, then in both cases momenta are necessarily commutative.

\subsection{GUP supercoordinates $X^i_0$ from canonical variables $(x,p)$}
\label{from_canonical_standard_to_GUP_super_subection}

Our algebra is given by (\ref{Super Noncommutative}). The Jacobi identity of interest is
\begin{equation}\label{Jacobi}
[X_0^i,[X_0^j,P_0^k]] +  [X_0^j,[P_0^k,X_0^i]] + [P_0^k,[X_0^i,X_0^j]] \equiv (\alpha) + (\beta) + (\gamma) =  0 \,,
\end{equation}
therefore
\begin{equation}\label{fromJacobi}
  [P_0^k,[X_0^i,X_0^j]] = - ((\alpha) + (\beta)) = {F^{kj}}_{, l} F^{il} - {F^{ki}}_{, l} F^{jl} \,,
\end{equation}
where ${F^{ij}}_{, k} = \partial F^{ij} / \partial P_0^k$. Let us compute $ - ((\alpha) + (\beta))$.

The right-hand side of (\ref{fromJacobi}) is antisymmetric in $i, j$, so it is identically zero for $i=j$. Now, consider the four cases left when $i \neq j$:
(a) $k=1 \,, j=1 \,, i=2$; (b) $k=1 \,, j=2 \,, i=1$; (c) $k=2 \,, j=1 \,, i=2$; d) $k=2 \,, j=2 \,, i=1$.

We have done explicitly all four computations, but given the antisymmetry, one can perform just one computation for $k=1$, say (a) and just one computation for $k=2$, say (c). We present that here. For (a)
\begin{equation}\label{fromJacobi a)}
  {F^{1 1}}_{, 1} F^{2 1} + {F^{1 1}}_{, 2} F^{2 2} - {F^{1 2}}_{, 1} F^{1 1} - {F^{1 2}}_{, 2} F^{1 2} = 0 \,,
\end{equation}
and for (c)
\begin{equation}\label{fromJacobi c)}
  {F^{2 1}}_{, 1} F^{2 1} + {F^{2 1}}_{, 2} F^{2 2} - {F^{2 2}}_{, 1} F^{1 1} - {F^{2 2}}_{, 2} F^{1 2} = 0 \,.
\end{equation}

Therefore
\begin{equation}\label{PXX is zero}
  [P_0^k,[X_0^i,X_0^j]] = 0 \,,
\end{equation}
which means, either $[X_0^i,X_0^j] = 0$ or $[X_0^i,X_0^j] = H^{i j} (\vec{P_0})$, where $H^{i j}$ is an arbitrary, antisymmetric in $i,j$, function of $\vec{P_0}$. Up to the order $O(\ell^2)$, the function  $H^{i j}$ must be proportional to the $\ell^2\epsilon^{ij}$, where $\epsilon^{ij}$ is the totally antisymmetric tensor in two dimensions\footnote{This is the only possibility compatible with our requirements, apart from $H^{i j}=0$. If one wants a result proportional to $P$, this requires a factor of $\ell^3$ to compensate for the physical dimension, which is bigger than $O(\ell^{2})$. If we include $\ell/P$, we do have the correct physical dimension, but $H^{ij}$ would not have a reasonable limit for $|\vec{P}_{0}|\to0$.}. This type of noncommutativity is known in the literature as the \textit{canonical noncommutativity} (see for instance \cite{Madore2000}).

The most natural choice for our algebra (\ref{Super Noncommutative}) is then
\begin{equation}\label{Super Noncommutative Snyder}
  [X_0^i , P_0^j] = \ii F^{i j} (\vec{P_0}) , \ \ \ [X_0^i , X_0^j] =  0 = [P_0^i , P_0^j]  \,,
\end{equation}
with
\begin{equation}\label{X0(x)}
  X_0^i = x^i \,,
\end{equation}
$P_0^i (\vec{p})$ given in (\ref{SupermomentumCartesian}) and $F^{i j} (\vec{P_0})$ given by (\ref{Fij(P0)}).

This is our first new result. We have found a GUP algebra already at the second layer, $(X_0, P_0)$, and we have discovered that we can actually use the standard (first layer) coordinates $x_i$ as the supercoordinates $X_0^i$. None of this was realized in \cite{ip2}. Now, together with $P_0^i (\vec{p})$ of (\ref{SupermomentumCartesian}), we have all the ingredients to study on graphene Dirac-like quantum dynamics in the presence of GUP corrections.

Notice that the trigonal warping breaks Lorentz (rotational here) symmetry, as can be seen by the structure of $F^{ij}$. Henceforth this is an algebra of the kind shown in (\ref{ADV comm rel}), where the GUP-induced terms do break Lorentz covariance. The only difference is in the shape of the $F^{ij}$s, which do not fully coincide with that expression. (The latter comes from phenomenology of QG). On the other hand, apparently the results discussed in \cite{AliDasVagenas1,*AliDasVagenas2,*AliDasVagenas3,*AliDasVagenas4,*AliDasVagenas5} do not depend on the exact expression of $F^{ij}$, as long as a term $O(\ell)$ is present there. Our expression complies with that, and actually also has $O(\ell^2)$ terms.

What is happening here (and in \cite{AliDasVagenas1,*AliDasVagenas2,*AliDasVagenas3,*AliDasVagenas4,*AliDasVagenas5}) is the following. When we move up in energy, i.e. when we consider contributions of the order $O(\ell^2)$ (but do not consider next-to-near neighbors, $A=0$) the supercoordinates are the same as the commutative coordinates, but the momenta become different. The uncertainty principle descending from $ [X_0^i , P_0^j]$ is then of a generalized kind but coordinates are of the standard, commutative kind.

In our case, though, when we include next-to-near neighbors we go one level up, to the hypervariables, $(X,P)$. As explained earlier, we better do so if we want to include all the $O(\ell^2)$ physics.

\subsection{Canonical supercoordinates $X^i_0$ from canonical variables $(x,p)$}

In this case we are looking for a canonical transformation, from the canonical variables $(x,p)$ to the canonical supervariables $(X_0,P_0)$, i.e. such that the algebra (\ref{Super Canonical}) is satisfied
\[
  [X_0^i , P_0^j] = \ii \delta^{i j}  , \ \ \ [X_0^i , X_0^j] = 0 = [P_0^i , P_0^j]   \,.
\]
This is achieved when
\begin{align}\label{supercoordinates_canonical}
X_{0}^{1} &= \left[1 + \frac{\ell}{2} p^{1} + \frac{\ell^{2}}{8}\left( 5 (p^{1})^{2} + 3 (p^{2})^{2} \right)\right] \, x^{1} + \left[-\frac{\ell}{2} p^{2} + \frac{\ell^{2}}{4} p^{1} p^{2}\right] \, x^{2}  \;, \nonumber \\
X_{0}^{2} &= \left[-\frac{\ell}{2} p^{2} + \frac{\ell^{2}}{4} p^{1} p^{2} \right] \, x^{1} + \left[1 - \frac{\ell}{2} p^{1} + \frac{\ell^{2}}{8}\left( 3 (p^{1})^{2} + 5 (p^{2})^{2} \right) \right] \, x^{2}  \;.
\end{align}
These, together with the expression (\ref{SupermomentumCartesian}) for $P^i_0(\vec{p})$
\begin{align*}
P^1_0 & =  p^1 + \frac{\ell}{4} \left((p^2)^2 - (p^1)^2\right) -  \frac{\ell^2}{8} p^1 \left((p^1)^2 + (p^2)^2\right) \;,  \\
P^2_0 & =  p^2 + \frac{\ell}{2} p^1 p^2 -  \frac{\ell^2}{8} p^2 \left((p^1)^2 + (p^2)^2\right)  \,,
\end{align*}
are the canonical transformation we were looking for. We note that it is possible to add an arbitrary function of momenta to the canonical supervariables $X_0$ in  (\ref{supercoordinates_canonical}) without spoiling the canonical structure. This freedom is related to the ordering ambiguity and does not affect other results.

We can say that (\ref{supercoordinates_canonical}) and (\ref{SupermomentumCartesian}) is a canonical transformation, as we are just changing one set of canonical variables $(\vec{X_{0}},\vec{P_{0}})$ to another one $(\vec{X},\vec{P})$ \cite{Goldstein}. Therefore, it is possible to change both coordinates and momenta, to keep the Poisson brackets unchanged, obtaining the same dynamical system. However, as seen from (\ref{supercoordinates_canonical}), this coordinate transformation is very ``unnatural'' from an experimental point of view [these $X_0^i (x,p)$ are not as simple and natural as $X_0^i = x^i$] and difficult to interpret. Nonetheless, it is remarkable that we found such an expression, as we have the choice to access a higher energy physics without altering the dynamical structure.

\section{From second to third layer}
\label{Sect_from_second_to_third_layer}

Upon introduction of the hypermomenta $P^i$ in \Cref{hypermomenta},
\[
P^{i}=P_{0}^{i}(1-A \, |\vec{P_{0}}|)\;,
\]
which, in turn, through (\ref{SupermomentumCartesian}) can be expressed in terms of standard variables $(x,p)$ as
\begin{align}\label{HypermomentaCartesian}
P^{1} & = p^{1} (1-A\,|\vec{p}|) + \ell \left(\frac{A\, (p^{1})^2 \left((p^{1})^{2} -3 (p^{2})^2\right)}{4 p}+\frac{1}{4} (1-A \,|\vec{p}|) \left((p^{2})^2-(p^{1})^2\right)\right)-\frac{1}{8} \,\ell^2 \,|\vec{p}|^2\, p^{1} \;, \\
P^{2} &= p^{2} (1-A \,|\vec{p}|) + \frac{\ell\, p^{2} \left(-A (p^{1})^3 - 5 A p^{1} (p^{2})^2+2 |\vec{p}| p^{1}\right)}{4 |\vec{p}|}-\frac{1}{8} \,\ell^2 \,|\vec{p}|^2 p^{2} \;, \nonumber
\end{align}
we have here the very same questions we had one layer below, that is, to find suitable hypercoordinates $X^i (X_0,P_0)$. Again there are many choices, and again we shall focus only on the most natural. At this point, given the previous results on the supervariables, we have \textit{three} natural choices: (i) GUP hypervariables from the canonical supervariables (\ref{Super Canonical}); (ii) Canonical hypervariables from the canonical supervariables (\ref{Super Canonical}); (iii) Unknown hypervariables from the GUP supervariables (\ref{Super Noncommutative Snyder}).

Let us proceed, case by case.

\subsection{Hypercoordinates $X_i$ from canonical supervariables $(X_{0},P_{0})$}

\subsubsection{GUP hypercoordinates $X^i$}
\label{from_canonical_super_to_GUP_hyper_subection}

Starting from the hypermomenta in terms of supermomenta (\ref{hypermomenta}),
and assuming the supercoordinates are canonical, $[X_{0}^{i},P_{0}^{j}]= i \delta^{ij}$, we get
\begin{equation}\label{commutators_mixing}
[X_{0}^{i},P^{j}]= \ii \left[ \delta^{ij} - A\, \left( |\vec{P_{0}}| \delta^{ij} + \frac{P_{0}^{i}P_{0}^{j}}{|\vec{P_{0}}|} \right) \right]\;.
\end{equation}

Once we are in the hypervariable layer, we should require
\begin{equation}\label{X(X0,P0)}
  X^i_0 \to X^i (X_0,P_0) \,,
\end{equation}
and to express the right-hand side of (\ref{commutators_mixing}) in terms of hypermomenta. To do so, we should invert the relation (\ref{hypermomenta}), which is not an exact expression but a series expansion in $A$. Indeed, see details in Appendix \ref{GUP_hyper_from_canonical_super_appendix},
\begin{align}\label{super_to_hypermomenta}
P_{0}^{i} & = P^{i}\,(1 + A\,|\vec{P}| + 2A^{2}\,|\vec{P}|^{2})  + \mbox{O}(A^{3}) \;, \quad \mbox{and} \nonumber \\
1-A\,|\vec{P}_{0}| & = 1-A\,|\vec{P}|-A^{2}|\vec{P}|^{2}  +  \mbox{O}(A^{3}) \;.
\end{align}
Taking into account all of this, we can write (\ref{commutators_mixing}) just as hypervariables,
\begin{equation}\label{commutators_hypervariables}
[X^{i},P^{j}]= \ii\,\left((1-A\,|\vec{P}|-A^{2}\,|\vec{P}|^{2})\delta^{ij}-A\,\frac{P^{i}P^{j}}{|\vec{P}|}\,(1+A\,|\vec{P}|)\right)\;.
\end{equation}
We demand the Jacobi identity to the commutator $[X^{i},X^{j}]$, i.e.,
\begin{equation}\label{Jacobi_X_X_P}
[X^{i},[X^{j},P^{k}]] +  [X^{j},[P^{k},X^{i}]] + [P^{k},[X^{i},X^{j}]] =  0 \,.
\end{equation}
Rearranging the last expression a little bit, we get
\begin{equation}
[P^{k},[X^{i},X^{j}]] = - [X^{i},[X^{j},P^{k}]] + [X^{j},[X^{i},P^{k}]] = - Y^{ijk} + Y^{jik} \;,
\end{equation}
where we defined $Y^{ijk}\equiv[X^{i},[X^{j},P^{k}]]$ in the last equality.

After some algebraic manipulations, more details in Appendix \ref{GUP_hyper_from_canonical_super_appendix},
\begin{equation*}
Y^{ijk} = A\,\frac{P^{i}}{|\vec{P}|}\,\delta^{jk} + A\,\frac{P^{k}}{|\vec{P}|}\,\delta^{ij} + A\,\frac{P^{j}}{|\vec{P}|}\,\delta^{ik}-A\,\frac{P^{j}P^{k}P^{i}}{|\vec{P}|^{3}} + \mbox{O}(A^{3}) \;.
\end{equation*}

We observe that $Y^{ijk}$ is totally symmetric in the three indices. Therefore,
\begin{equation*}
[P^{k},[X^{i},X^{j}]] = - Y^{ijk} + Y^{jik} = 0 \;,
\end{equation*}
up to order $O(A^{2})$. This means, either $[X^{i},X^{j}]=0$ or $[X^{i},X^{j}] = H^{i j} (P)$, where $H^{i j}$ is an arbitrary, antisymmetric in $(i,j)$ indices, function of $P$.  Again, the most natural choice for our algebra is
{\footnotesize\begin{equation}\label{Hyper_commutative_GUP}
  [X^{i} , P^{j}] = \ii\left[ \delta^{ij} - A\, |\vec{P}|\,\left(  \delta^{ij} + \frac{P^{i}P^{j}}{|\vec{P}|^{2}} \right)
  - A^{2} \,|\vec{P}|^{2}\,\left( \delta^{ij} + \frac{P^{i}P^{j}}{|\vec{P}|^{2}} \right) \right] , \ \ \ [X^{i} , X^{j}] =  0  , \ \ \  [P^{i} , P^{j}] = 0  \;.
\end{equation}}
This is very similar to the expression (\ref{ADV comm rel}), emerging from QG, which we have discussed earlier in the Introduction, including the sign of the terms proportional to $A$, but not the sign nor all the coefficients of the terms proportional to $A^{2}$. Let us notice, that such $O(A^2)$ terms are not those descending from considering the $m=3$ contributions, i.e., the next-to-next-to-near neighbors (we may as well do that, but we would not keep any Dirac Hamiltonian structure, not even deformed). The $O(A^2)$ terms we have here are simply those we have obtained by manipulating the $O(A)$ terms, as shown above. We have to do so, because we are actually doing $O(\ell^2)$ calculations.

This is another new result of this paper, which improves the findings of \cite{ip2}. Indeed, there $O(A^2)$ terms were not included at all. Another important result we obtain here is that the hypercoordinates $X^i$ can be expressed in terms of the supercoordinates $X_0^i$ as
\begin{equation}\label{Xi}
X^{i}=X_{0}^{i} \;,
\end{equation}
so that we have full control of all the phase-space hypervariables.

By substituting (\ref{supercoordinates_canonical}) in (\ref{Xi}), we get
\begin{align*}
X^{1} &= \left[1 + \frac{\ell}{2} p^{1} + \frac{\ell^{2}}{8}\left( 5 (p^{1})^{2} + 3 (p^{2})^{2} \right)\right] \, x^{1} + \left[-\frac{\ell}{2} p^{2} + \frac{\ell^{2}}{4} p^{1} p^{2}\right] \, x^{2}  \;,  \\
X^{2} &= \left[-\frac{\ell}{2} p^{2} + \frac{\ell^{2}}{4} p^{1} p^{2} \right] \, x^{1} + \left[1 - \frac{\ell}{2} p^{1} + \frac{\ell^{2}}{8}\left( 3 (p^{1})^{2} + 5 (p^{2})^{2} \right) \right] \, x^{2}  \;,
\end{align*}
which along with (\ref{HypermomentaCartesian}) give us the canonical transformation from standard variables $(x,p)$ to canonical hypervariables $(X,P)$.

\subsubsection{Canonical hypercoordinates $X^i$}

The other natural choice is to define the hypercoordinates $X^i$ in terms of the supervariables $X^i_{0}$ and $P^i_{0}$ in such a way that they satisfy the canonical Heisenberg algebra
\begin{equation}\label{commutators2}
[X^{i},P^{j}]= \ii \delta_{ij}\,, \; [X^{i},X^{j}]=0 \,, \; [P^{i},P^{j}]=0 \;.
\end{equation}
In this case, the hypercoordinates $X^{i}$ are (see details in Appendix \ref{canonical_hyper_from_canonical_super_appendix})
\begin{equation}\label{first_relation}
X^{i} = \frac{X^{i}_{0}}{1-A\,|\vec{P_{0}}|} +  \frac{A (X^j_{0}P_{0 j}) P^{i}_{0}}{|\vec{P_{0}}|\,(1-A\,|\vec{P_{0}}|)(1-2A\,|\vec{P_{0}}|)} \;.
\end{equation}
This is what done, to a certain extent, in \cite{ip2} (see Section 3 on that paper).

We can write the hypercoordinates $\vec{X}$ of this case in terms of the standard variables $\vec{x}$ and $\vec{p}$. Indeed, after substituting (\ref{supercoordinates_canonical}) and (\ref{SupermomentumCartesian}) in (\ref{first_relation}) we obtain
\begin{align}\label{hypercoordinates_canonical_GUP_super}
X^{1} &= x^{1}+\frac{1}{2} \,\ell\, (p^{1} x^{1}-p^{2} x^{2})+ \frac{1}{8} \,\ell^2 \left(5 (p^{1})^2 x^{1}+2 p^{1} p^{2} x^{2}+3 (p^{2})^2 x^{1}\right)\\
&+\frac{A \left(2 (p^{1})^2 x^{1}+p^{1} p^{2} x^{2}+(p^{2})^2 x^{1}\right)}{|\vec{p}|}+A^2 \left(4 (p^{1})^2 x^{1}+3 p^{1} p^{2} x^{2}+(p^{2})^2 x^{1}\right) \nonumber\\
&+\frac{A \,\ell\, \left(2 (p^{1})^5 x^{1}-4 (p^{1})^4 p^{2} x^{2}+3 (p^{1})^3 (p^{2})^2 x^{1}-9 (p^{1})^2 (p^{2})^3 x^{2}+5
   p^{1} (p^{2})^4 x^{1}-(p^{2})^5 x^{2}\right)}{4 |\vec{p}|^{3}} \nonumber\\
X^{2} &= x^{2} - \frac{1}{2} \,\ell\, (p^{1}\, x^{2}+p^{2}\,x^{1})
+\frac{1}{8} \,\ell^2 \left(3 (p^{1})^2 x^{2}+2 p^{1}\, p^{2} x^{1}+5 (p^{2})^2 x^{2}\right) \;, \nonumber \\
+&\frac{A \left((p^{1})^2 x^{2}+p^{1}\, p^{2} x^{1}+2 (p^{2})^2 x^{2}\right)}{|\vec{p}|} + A^2 \left((p^{1})^2 x^{2}+3 p^{1}\, p^{2} x^{1}+4 (p^{2})^2 x^{2}\right) \nonumber\\
-&\frac{A \,\ell\, \left(3 (p^{1})^5 x^{2}-2 (p^{1})^4 p^{2} x^{1}+(p^{1})^3 (p^{2})^2 x^{2}+5 (p^{1})^2 (p^{2})^3 x^{1}+2
   p^{1}\, (p^{2})^4 x^{2}+3 (p^{2})^5 x^{1}\right)}{4 |\vec{p}|^{3}} \;. \nonumber
\end{align}
Here, (\ref{hypercoordinates_canonical_GUP_super}) and (\ref{HypermomentaCartesian}) are the canonical transformation from standard variables $(x,p)$ to canonical hypervariables $(X,P)$.

\subsection{Hypercoordinates $X_i$ from GUP supervariables $(X_{0},P_{0})$}

The GUP supervariables are found to be $X_0^i = x^i$ and $P_0^i (p)$ in (\ref{SupermomentumCartesian}). Knowing that the hypermomentum $P^i (P_0)$ is given by (\ref{hypermomenta}), we want to find now $X^{i} (X_{0}, P_{0})$.

\subsubsection{Canonical hypercoordinates $X^i$}

Let us suppose the supervariables have the GUP algebra (\ref{Super Noncommutative Snyder}). We shall find the relation $X^{i} (X_{0}, P_{0})$ such that the hypervariables are canonical, i.e., $[X^{i},P^{j}]= \ii\,\delta^{ij}$ and $[X^{i},X^{j}]=0=[P^{i},P^{j}]$.

Let us take the ansatz
\begin{equation}\label{hypercoordinates_GUP_supercoordinates}
X^{i}=G^{ij}\,X_{0}^{j} \;,
\end{equation}
where $G^{ij}=G^{ij}(A\,P_{0})$ is a dimensionless function. Therefore,
\begin{equation*}
\begin{split}
  [X^{i},P^{j}] &= [G^{ik}\,X_{0}^{k},P_{0}^{j}(1-A\,|\vec{P_{0}}|)] \\
                &= G^{ik}\,[X_{0}^{k},P_{0}^{j}](1-A\,|\vec{P_{0}}|) - A\,G^{ik}\,P_{0}^{j}\,[X_{0}^{k},|\vec{P_{0}}|] \\
                &= \ii\,G^{ik}\,F^{kj}\,(1-A\,|\vec{P_{0}}|) - A\,G^{ik}\,P_{0}^{j}\,\frac{P_{0}^{l}}{|\vec{P_{0}}|}\,F^{lk} \\
                &= \ii\,G^{ik}\left(F^{kj}\,(1-A\,|\vec{P_{0}}|) - A\,P_{0}^{j}\,\frac{P_{0}^{l}}{|\vec{P_{0}}|}\,F^{lk}\right) \\
                &= \ii\,\delta^{ij} \;.
\end{split}
\end{equation*}
So, $G^{ij}$ should be the inverse of the matrix
\begin{equation}\label{K_definition}
K^{ij} = F^{ij}\,(1-A\,|\vec{P_{0}}|) - A\,P_{0}^{j}\,\frac{P_{0}^{k}}{|\vec{P_{0}}|}\,F^{ki} \;.
\end{equation}
The result up to order $O(\ell^{2})$ is
{\footnotesize \begin{equation}\label{G functions}
\begin{split}
G^{11} &= 1 + \frac{\ell}{2}\,(P_{0}^{1} + 2A\,\frac{(P_{0}^{1})^{3}}{|\vec{P_{0}}|}) + A\,\frac{(2(P_{0}^{1})^{2} + (P_{0}^{2})^{2})}{|\vec{P_{0}}|}  + \frac{\ell^{2}}{4}\, (3(P_{0}^{1})^{2}+(P_{0}^{2})^{2}) + A^{2}\,(4(P_{0}^{1})^{2}+(P_{0}^{2})^{2}) \;,  \\
G^{12} &= -\frac{\ell}{2}\,(P_{0}^{2} + A\,\frac{P_{0}^{2}\,(3(P_{0}^{1})^{2}+(P_{0}^{2})^{2})}{|\vec{P_{0}}|}) + A\,\frac{P_{0}^{1}\,P_{0}^{2}}{|\vec{P_{0}}|} + \frac{\ell^{2}}{2}\,P_{0}^{1}\,P_{0}^{2} + 3A^{2}\,P_{0}^{1}\,P_{0}^{2}  \;,  \\
G^{21} &= -\frac{\ell}{2}\,(P_{0}^{2} + A\,\frac{(P_{0}^{2})^{3}}{|\vec{P_{0}}|}) + A\,\frac{P_{0}^{1}\,P_{0}^{2}}{|\vec{P_{0}}|} + \frac{\ell^{2}}{2}\,P_{0}^{1}\,P_{0}^{2} + 3A^{2}\,P_{0}^{1}\,P_{0}^{2}\;, \\
G^{22} &= 1 - \frac{\ell}{2}\,(P_{0}^{1} + A\,P_{0}^{1}\,\frac{(P_{0}^{1})^{2}+3(P_{0}^{2})^{2}}{|\vec{P_{0}}|}) + A\,\frac{((P_{0}^{1})^{2} + 2(P_{0}^{2})^{2})}{|\vec{P_{0}}|}  + \frac{\ell^{2}}{4}\, ((P_{0}^{1})^{2}+3(P_{0}^{2})^{2}) + A^{2}\,((P_{0}^{1})^{2}+4(P_{0}^{2})^{2})\;.
\end{split}
\end{equation}}
Using (\ref{SupermomentumCartesian}) and (\ref{X0(x)}) we can express the hypercoordinates $X^{i}$ in terms of standard variables $(x,p)$. Indeed,
\begin{equation}\label{X(x)}
\begin{split}
X^{1} &= G^{11}(\vec{p})\,x^{1} + G^{12}(p)\,x^{2} \;,  \\
X^{2} &= G^{21}(\vec{p})\,x^{1} + G^{22}(p)\,x^{2} \;,
\end{split}
\end{equation}
where $G^{ij}(\vec{p})$ are the entries \eqref{G functions} of the matrix $G$ expressed in terms of $\vec{p}$ through \eqref{SupermomentumCartesian}. At order $O(\ell^{2})$,
{\footnotesize \begin{equation} \label{G functions_x}
\begin{split}
G^{11}(\vec{p}) &= 1 + \frac{\ell}{2}\,(p^{1} + A\,p^{1}\,\frac{6(p^{1})^{2}-((p^{2})^{2}}{2|\vec{p}|}) + A\,\frac{(2(p^{1})^{2} + (p^{2})^{2})}{|\vec{p}|}  + \frac{\ell^{2}}{8}\, (7(p^{1})^{2}+(p^{2})^{2}) + A^{2}\,(4(p^{1})^{2}+(p^{2})^{2}) \;,  \\
G^{12}(\vec{p}) &= -\frac{\ell}{2}\,(p^{2} + A\,\frac{p^{2}\,(4(p^{1})^{2}+3(p^{2})^{2})}{2|\vec{p}|}) + A\,\frac{p^{1}\,p^{2}}{|\vec{p}|} + \frac{\ell^{2}}{4}\,p^{1}\,p^{2} + 3A^{2}\,p^{1}\,p^{2}  \;,  \\
G^{21}(\vec{p}) &= -\frac{\ell}{2}\,(p^{2} + A\,\frac{p^{2}\,(2(p^{1})^{2}-5(p^{2})^{2})}{2|\vec{p}|}) + A\,\frac{p^{1}\,p^{2}}{|\vec{p}|} + \frac{\ell^{2}}{4}\,p^{1}\,p^{2} + 3A^{2}\,p^{1}\,p^{2}\;, \\
G^{22}(\vec{p}) &= 1 - \frac{\ell}{2}\,(p^{1} + A\,p^{1}\,\frac{(p^{1})^{2}+2(p^{2})^{2}}{2|\vec{p}|}) + A\,\frac{((p^{1})^{2} + 2(p^{2})^{2})}{|\vec{p}|}  + \frac{\ell^{2}}{8}\, ((p^{1})^{2}+7(p^{2})^{2}) + A^{2}\,((p^{1})^{2}+4(p^{2})^{2})\;.
\end{split}
\end{equation}}
Now, it is worth to mentioning once this expression is substituted in transformation (\ref{X(x)}), after some rearrangements we get exactly the same transformation (\ref{hypercoordinates_canonical_GUP_super}). This is a reassuring result, as it tells us that going from standard variables $(x,p)$ to canonical hypervariables $(X,P)$ is the same canonical transformation, it does not matter whether we pass through canonical or GUP supervariables $(X_{0},P_{0})$.

\subsubsection{GUP hypercoordinates $X^i$}

Let us suppose again the supervariables have the GUP algebra (\ref{Super Noncommutative Snyder}), but now we allow that hypervariables are not necessarily canonical. Proceeding with the same reasoning line of Section \ref{Sect_from_first_to_second_layer}, we have
\begin{equation*}
[X_{0}^{i},P^{j}] = [X_{0}^{i},P_{0}^{j}(1-A\,|\vec{P_{0}}|)] = \ii\,F^{i\,k}(\vec{P_{0}})\,\left((1-A\,|\vec{P_{0}}|)\delta^{kj}-A\,\frac{P_{0}^{k}\,P_{0}^{j}}{|\vec{P_{0}}|}\right) \;.
\end{equation*}
We want to express $F^{i\,j}$ as a function of hypermomenta $\vec{P}$. By using (\ref{super_to_hypermomenta}), we get
\begin{equation*}
[X_{0}^{i},P^{j}] = \ii\,F^{i\,k}(\vec{P})\,\left((1-A\,|\vec{P}|-A^{2}\,|\vec{P}|^{2})\delta^{kj}-A\,\frac{P^{k}\,P^{j}}{|\vec{P}|}\,(1+A\,|\vec{P}|)\right) \;.
\end{equation*}
Taking inspiration from the earlier sections, we can ask if it is allowed to take the choice for the algebra of hypervariables
\begin{equation}\label{hyper_algebra}
[X^{i},P^{j}] =  \ii\,\mathcal{F}^{i\,j}(\vec{P}) \;, \ \  [X^{i},X^{j}] = \mathcal{G}^{i\,j}  \;, \ \   [P^{i},P^{j}] = 0 \;,
\end{equation}
with
\begin{equation} \label{CallFij}
\mathcal{F}^{i\,j}(\vec{P}) \equiv F^{i\,k}(\vec{P})\,\left((1-A\,|\vec{P}|-A^{2}\,|\vec{P}|^{2})\delta^{kj}-A\,\frac{P^{k}\,P^{j}}{|\vec{P}|}\,(1+A\,|\vec{P}|)\right) \;,
\end{equation}
for the case the supercoordinates are the same as the hypercoordinates, i.e.,
\begin{equation}\label{canonical_hyper_from_canonical_super}
  X^{i} = X_{0}^{i} \;.
\end{equation}
This last would be true if the Jacobi identity (\ref{Jacobi_X_X_P}) holds for the option (\ref{hyper_algebra}) and (\ref{canonical_hyper_from_canonical_super}). It can be checked this is indeed the case (see Appendix \ref{GUP_hyper_from_GUP_super_appendix}) for the case $\mathcal{G}^{i\,j}$ is an arbitrary, antisymmetric in $(i,j)$ indices, function of $P$. Once more, the most natural choice for our algebra is
\begin{equation}\label{hyper_GUP}
[X^{i},P^{j}] =  \ii\,\mathcal{F}^{i\,j}(\vec{P}) \;, \ \ \ [X^{i},X^{j}] = 0  \;, \ \ \  [P^{i},P^{j}] = 0 \;.
\end{equation}
It can be checked this is a GUP that somehow extends the other two GUPs found in \cref{from_canonical_standard_to_GUP_super_subection,from_canonical_super_to_GUP_hyper_subection}. Indeed, by taking into account only near neighbors ($A=0$) then $F^{i\,j}(\vec{P})=\mathcal{F}^{i\,j}(\vec{P})$ and we get exactly (\ref{Super Noncommutative Snyder}) once we express $\vec{P}$ in terms of $\vec{P_{0}}$. On the other hand, starting from canonical supervariables ($F^{i\,j}=\delta^{ij}$) and then (\ref{Hyper_commutative_GUP}) raises.
It is also important to mention that in this general case
\begin{equation*}
  X^{i} = X_{0}^{i} = x^{i} \;,
\end{equation*}
meaning that supercoordinates and hypercoordinates both coincide with the standard coordinates.

\section{Conclusions}

We have clearly identified the three sets of natural variables that mimic three different levels of high-energy scenarios that we call ``world'', $(x,p)$, ``super-world'', $(X_0,P_0)$, and ``hyper-world'', $(X,P)$. Starting from the ``world'', which refers to variables we routinely measure in the lab, and hence are canonical, we found maps to the next two levels where three different GUPs naturally appear.

Let us repeat here what we have done to summarize the above and to give in one place the key facts and formulae. We first moved from the standard, canonical variables that describe the low energy physics of the Dirac quasiparticles of graphene, $(x,p)$, to the next level that we called supervariables, $(X_0,P_0)$. Then we moved one more ``layer'' up, to the hypervariables $(X,P)$, which means to include the next-to-near neighbors in the computations ($A \neq 0$). The number of options grows according to the following diagram
\begin{equation*}
\begin{tikzcd}[column sep=huge]
                                                   &                                                       & \mbox{canonical $(X,P)$ $(IV)$}  \\
                                                   &   \mbox{canonical $(X_0,P_0)$ $(II)$} \ar[ur] \ar[dr] &                                  \\
                                                   &                                                       & \mbox{GUP $(X,P)$ $(V)$}         \\
\mbox{canonical $(x,p)$ $(I)$}  \ar[uur] \ar[ddr]  &                                                       &                                  \\
                                                   &                                                       & \mbox{canonical $(X,P)$ $(VI)$}  \\
                                                   &   \mbox{GUP $(X_0,P_0)$ $(III)$} \ar[ur] \ar[dr]      &                                  \\
                                                   &                                                       & \mbox{GUP $(X,P)$ $(VII)$}
\end{tikzcd}
\end{equation*}

Particularly important seems to us the ladder that goes from the lab variables all the way up to GUPs in both higher levels, that is, the ladder $(I) \to (III) \to (VII)$. Indeed, this ladder is obtained with the simple identifications
\[
x^i = X_0^i = X^i \,,
\]
that means that the original lab coordinates, $x^i$, along with the generalized momenta, $P_0^i(p)$ and $P^i(p)$, are natural coordinates to describe rich GUP physics. This is very encouraging because nothing needs to be done on such coordinates for the system to reproduce important GUPs, otherwise very challenging to grasp in direct (Planck length) experiments.

In going from $(I)$ to $(III)$ we have the first GUP, that of Eqs. (\ref{Super Noncommutative Snyder}),
\[
  [X_0^i , P_0^j] = i F^{i j} (\vec{P_0})  \,,
\]
and $[X_0^i , X_0^j] =  0 = [P_0^i , P_0^j]$, with $F^{i j} (\vec{P_0})$ given in (\ref{Fij(P0)}),
\[
  F^{i j} (\vec{P_0}) = \delta^{i j}
  + \frac{1}{2} \, \ell \, \left(
                  \begin{array}{cc}
                  - P^1_0 & P^2_0 \\
                    P^2_0 & P^1_0 \\
                  \end{array}
                \right)
  - \frac{1}{2} \, \ell^2 \left(
                  \begin{array}{cc}
                    (P^1_0)^2  & P^1_0 P^2_0 \\
                    P^1_0 P^2_0 &  (P^2_0)^2   \\
                  \end{array}
                \right) \,,
\]
which is an unusual GUP, but surely an easy one to spot. Then, proceeding from $(III)$ to $(VII)$, we obtain the beautiful GUP of Eqs. (\ref{hyper_GUP}),
\[
[X^{i},P^{j}] =  \ii \,\mathcal{F}^{i\,j}(\vec{P})
\equiv \ii F^{i\,k}(\vec{P})\,\left[ \delta^{kj} - A\, |\vec{P}|\,\left(  \delta^{kj} + \frac{P^{k}P^{j}}{|\vec{P}|^{2}} \right)
  - A^{2} \,|\vec{P}|^{2}\,\left( \delta^{kj} + \frac{P^{k}P^{j}}{|\vec{P}|^{2}} \right) \right]
\]
where $[X^{i},X^{j}] = 0  = [P^{i},P^{j}]$. As we shall comment in a moment, this GUP contains the other two GUPs and generalizes them.

Also noticeable is the ``canonical way,'' which is the ladder of the canonical transformations all the way from the first to the last level, $(I) \to (II) \to (IV)$. This is achieved by paying the price of complicated expressions for $X_0^i (x,p)$ and $X^i(x,p)$, which somehow suggest that the most natural way is the one including GUPs. In particular, the transformation from $(I)$ to $(II)$ is achieved when $X_0^i (x,p)$ is given by (\ref{supercoordinates_canonical})
\begin{align}
X_{0}^{1} &= \left[1 + \frac{\ell}{2} p^{1} + \frac{\ell^{2}}{8}\left( 5 (p^{1})^{2} + 3 (p^{2})^{2} \right)\right] \, x^{1} + \left[-\frac{\ell}{2} p^{2} + \frac{\ell^{2}}{4} p^{1} p^{2}\right] \, x^{2}  \;, \nonumber \\
X_{0}^{2} &= \left[-\frac{\ell}{2} p^{2} + \frac{\ell^{2}}{4} p^{1} p^{2} \right] \, x^{1} + \left[1 - \frac{\ell}{2} p^{1} + \frac{\ell^{2}}{8}\left( 3 (p^{1})^{2} + 5 (p^{2})^{2} \right) \right] \, x^{2}  \;, \nonumber
\end{align}
and clearly the momenta $P_0^i(p)$ are those of (\ref{SupermomentumCartesian}). To take the next step, that is from $(II)$ to $(IV)$, one needs coordinates as given in (\ref{first_relation}),
\[
X^{i} = \frac{X^{i}_{0}}{1-A\,|\vec{P_{0}}|} +  \frac{A (X^j_{0}P_{0 j}) P^{i}_{0}}{|\vec{P_{0}}|\,(1-A\,|\vec{P_{0}}|)(1-2A\,|\vec{P_{0}}|)} \;,
\]
which together with the momenta in (\ref{hypermomenta}) are the canonical transformation from the supervariables to the hypervariables.

We can go all the way from the standard variables, by expressing $X^i(x,p)$, as given in (\ref{hypercoordinates_canonical_GUP_super}), together with $P^i(p)$ as given in (\ref{HypermomentaCartesian}).

In between these two ladders (the canonical all-the-way and the GUP all-the-way) we also have two more ``middle-way'' ladders: $(I) \to (II) \to (V)$ and $(I) \to (III) \to (VI)$. The former ends with the GUP in the hypervariables given in (\ref{Hyper_commutative_GUP})
\[
  [X^{i} , P^{j}] = \ii\left[ \delta^{ij} - A\, |\vec{P}|\,\left(  \delta^{ij} + \frac{P^{i}P^{j}}{|\vec{P}|^{2}} \right)
  - A^{2} \,|\vec{P}|^{2}\,\left( \delta^{ij} + \frac{P^{i}P^{j}}{|\vec{P}|^{2}} \right) \right]  \,,
\]
and $[X^{i} , X^{j}] =  0 = [P^{i} , P^{j}]$ with $X^i = X_0^i (x,p)$ given in (\ref{supercoordinates_canonical}). The latter ends with canonical hypervariables given by $X^i(X_0,P_0)$ in (\ref{hypercoordinates_GUP_supercoordinates}) with (\ref{G functions}), and they can then be expressed in terms of the standard variables $(x,p)$, as done in (\ref{X(x)}) with (\ref{G functions_x}).

This summary should have made clear that, having set under control the mappings between the various layers, we have obtained here two more GUPs: the one in (\ref{Super Noncommutative Snyder}), $[X_0^i , P_0^j] = \ii \,F^{i j} (\vec{P_0})$, and the one in (\ref{hyper_GUP}), $[X^{i},P^{j}] =  \ii \,\mathcal{F}^{i\,j}(\vec{P})$, where $X_0^i = x^i$ and $X^i = x^i$, respectively. The other GUP of (\ref{Hyper_commutative_GUP}), reported in the previous paragraph, is an improvement of the one obtained earlier \cite{ip2} in two respects: we now have $X(x,p)$ and we now have $A^2$ terms. Notice that the latter GUP is very similar to the expression (1), that emerged from QG and that we discussed
in the Introduction. This similarity includes the sign of the terms proportional to $A$, but not the sign nor any of the coefficients of the terms proportional to $A^2$.

These results are all consistent. Indeed, on the one hand, $\mathcal{F}^{i\,j}(\vec{P}) \to F^{i\,j}(\vec{P_0})$, when $A = 0$, hence there is no ``hyperworld'' and the first GUP in the ``superworld'' is all we have. On the other hand, when the ``superworld'' is canonical, i.e.,  $F^{i\,j}(\vec{P_0}) \to \delta^{i j}$, then $\mathcal{F}^{i\,j}(\vec{P})$ reduces to the GUP of (\ref{Hyper_commutative_GUP}) just discussed.

To conclude, let us say that the results of this paper pave the way to the use of graphene, and other Dirac materials, as table-top laboratories to reproduce GUP-corrected scenarios. First, in two crucial cases with GUPs, the high energy coordinates just coincide with the low-energy, measurable coordinates. In the other cases, the high-energy coordinates have complicated expressions in terms of the standard, measurable phase-space variables, and are less appealing for experiments. Therefore, the most natural road to properly include the higher order effects due to the lattice structure appear to be the one that passes through the GUPs, what we indicated as the ladder $ (I) \to (III) \to (VII)$.

The high-energy expressions of the momenta, $\vec{P}_0$ and $\vec{P}$, are given once and for all in terms of the low-energy, measurable momenta, $\vec{p}$, by taking into account the effect of the dispersion relations at higher and higher order. In other words, it is the physics of graphene that dictates these expressions.

Let us stress here that each pair of variables $(x,p)$, $(X_{0},P_{0})$ or $(X,P)$ does not define by themselves the evolution of the system, as it is needed also their commutations algebra. This is why, for instance, the options $(IV)-(VII)$ are differentiated (also options $(II)$ and $(III)$), as even they are on the same ``layer'', but with different commutator relations.

Here we have all these expressions under control. The dynamics of the quasiparticles is given by the Hamiltonian (\ref{Hamiltonian_near_neighbors}), which means to stop at the $\vec{P}_0$, or by the Hamiltonian (\ref{Hamiltonian_total_Op3}), when $\vec{P}$ is considered. All of that is written in terms of the measurable $\vec{p}$, the only momenta we have access to. All we need to do, then, is to recognize the generalized momenta (and, when necessary, the generalized coordinates) in the given expression. When we do that, we reconstruct the dynamics as dictated by the Hamiltonians (\ref{newDiracHwithP0}) or (\ref{H(P)}), respectively, and this allows us to experimentally see the effects of the GUPs on the physics of the Dirac quasiparticles. This appears to us quite remarkable.

Indeed, one does not need any special or exotic setup to have the GUP all-the-way ladder at work. The only thing that is necessary is to include higher order terms in the dispersion relations, and then move to the next-to-near neighbors. The coordinates are always just $x^i$. Since the literature is full of proposals on how GUPs affect a variety of phenomena, see, e.g., \cite{Maggiore:1993rv,Kempf:1993bq,Scardigli:1999jh,Buoninfante:2019fwr,Petruzziello:2020wkd,Bosso:2021koi,GUPBTZ} and since we have obtained here many such GUPs, some very similar to the ones descending from QG, then we are now in the position to prove some of those theoretical conjectures in the lab.

We shall not present here any specific proposals for an experiment. Nonetheless, let us illustrate a recipe of how all the above could be implemented in an experiment. The most direct way to see the phase space variables at work in the lab, and the GUPs among them, is to have the Dirac quasi-particle propagating in a potential, say it
\[
V(X) \,,
\]
as done, for instance, in \cite{KleinGraphene}. This way we do not need to deal with other kinds of particles, other than the Dirac massless particles. They are the kind of particles that surely reflect the modified phase space variables, leading to GUPs or else. This is quite an important point, as other particles may or may not come from the outer world, with respect to the graphene membrane. In the case they come from outside (think of a photon from an external source), their space variables may not coincide with those of the ``graphene world'' \footnote{In fact, we may even think of engineering a very specific set-up for which the photon, although born in three space dimensions, only propagates in two dimensions, hence some sort of solution can be found. Nonetheless, as well known, a truly two-space dimensional photon differs in more substantial ways from a three-space dimensional photon projected in two dimensions (see, for instance, the works on \emph{reduced QED} \cite{Marino:1992xi,Gorbar:2001qt}). In other words, the best one can do here, to reproduce conditions that keep the experiment within the limits of the ``graphene world'', is that the particles/excitations/gauge fields are all ``born in that world'', like the Dirac quasi-particles. Some examples are discussed here \cite{Acquaviva:2022yiq}.}, that is, with the membrane space variables $X^i$ (or $X_0^i$, for that matters).

Having said that, let us consider first the case of GUPs all the way, that means $X^i=x^i$. In this case the potential well is simply $V(x)$, and everything goes through as in a standard (non GUP-corrected) case, except for the fact that the Hamiltonian now has higher order derivatives corrections. One needs to work-out the solutions to $H \Psi = E \Psi$, and consider the propagation in that potential. The result should be the GUP-corrected dynamics of the massless Dirac particle in that potential and could be compared with the experimental data.

Other cases may be treated, too, but less straightforwardly. For instance, the ``middle-way'' ladder $(I) \to (II) \to (V)$, which ends with the closest GUP to the one of \cite{AliDasVagenas1} given in (\ref{Hyper_commutative_GUP}),
is such that $X^i = X_0^i (x,p)$ as given in (\ref{supercoordinates_canonical}). This leads to a $V(x,p)$, which is certainly less easy to realize in practice than a simple $V(x)$, for instance, of the well type. Nonetheless, in principle, it is possible.

\section*{Acknowledgements}
A.~I. thanks the INFN, Sezione di Cagliari, for supporting a visit to Cagliari University, where parts of this work were carried on. A.~I. and P.~P. gladly acknowledge support from Charles University Research Center (UNCE/SCI/013). P.~P. is also supported by Fondo Nacional de Desarrollo Cient\'{i}fico y Tecnol\'{o}gico--Chile (Fondecyt Grant No.~3200725). S.~M. acknowledges support from COST action CA18108.

\appendix

\section{GUP hypervariables from canonical supervariables: more details}
\label{GUP_hyper_from_canonical_super_appendix}

Starting from the hypermomenta in terms of supermomenta \eqref{hypermomenta}, and assuming the supercoordinates are canonical, $[X_{0}^{i},P_{0}^{j}]=\ii \delta^{ij}$, we get the commutators \eqref{commutators_mixing}.

In the hypervariable layer, we should require
\begin{equation*}
  X^i_0 \to X^i (\vec{X}_0,\vec{P}_0) \,,
\end{equation*}
and to express the right-hand side of \eqref{commutators_mixing} in terms of hypermomenta. To do so, we should invert the relation \eqref{hypermomenta}, which is not an exact expression but a series expansion in $A$. Indeed,
\begin{equation*}
P_{0}^{i}=P^{i}\,(1-A\,|\vec{P_{0}}|)^{-1}= P^{i}\,(1+A\,|\vec{P_{0}}|+A^{2}\,|\vec{P_{0}}|^{2}) + \mbox{O}(A^{3})
\end{equation*}
As $|\vec{P}|\,\equiv\,(P^{i}P_{i})^{1/2}=(P_{0}^{i}P_{0\,i})^{1/2}\,(1-A\,|\vec{P_{0}}|)^{2/2}=|\vec{P_{0}}|\,(1-A\,|\vec{P_{0}}|)$, therefore,
\begin{equation}
|\vec{P_{0}}|=|\vec{P}|\,(1-A\,|\vec{P_{0}}|)^{-1}=|\vec{P}|\,(1+A\,|\vec{P_{0}}|+A^{2}\,|\vec{P_{0}}|^{2}) + \mbox{O}(A^{3})
\end{equation}
Proceeding in an iterative way, up to order $O(A^{2})$,
\begin{equation*}
|\vec{P_{0}}|=|\vec{P}|\,(1+A\,|\vec{P}|\,(1+A\,|\vec{P}|)+A^{2}\,|\vec{P}|^{2})  = P^{i}\,(1 + A\,|\vec{P}| + 2A^{2}\,|\vec{P}|^{2})\;,
\end{equation*}
which gives the expressions (\ref{super_to_hypermomenta}) of supermomenta in terms of hypermomenta, up to order $O(A^{2})$,
\begin{equation*}
P_{0}^{i} = P^{i}\,(1 + A\,|\vec{P}| + 2A^{2}\,|\vec{P}|^{2}) \;,
\end{equation*}
and
\begin{equation*}
1-A\,|\vec{P}_{0}| = 1-A\,|\vec{P}|-A^{2}|\vec{P}|^{2} \;.
\end{equation*}
We can now write \eqref{commutators_mixing} just as hypervariables,
\begin{equation}\label{commutators_hypervariables}
[X^{i},P^{j}]=\ii\left((1-A\,|\vec{P}|-A^{2}\,|\vec{P}|^{2})\delta^{ij}-A\,\frac{P^{i}P^{j}}{|\vec{P}|}\,(1+A\,|\vec{P}|)\right)\;.
\end{equation}

Let us infer how  the commutator $[X^{i},X^{j}]$ looks by demanding the Jacobi identity
\begin{equation*}
[X^{i},[X^{j},P^{k}]] +  [X^{j},[P^{k},X^{i}]] + [P^{k},[X^{i},X^{j}]] =  0 \,.
\end{equation*}
Rearranging the last expression, we get
\begin{equation*}
[P^{k},[X^{i},X^{j}]] = - [X^{i},[X^{j},P^{k}]] + [X^{j},[X^{i},P^{k}]] = - Y^{ijk} + Y^{jik} \;,
\end{equation*}
where we defined $Y^{ijk}\equiv[X^{i},[X^{j},P^{k}]]$ in the last equality.

Let us compute first $Y^{ijk}$ up to order $O(A^{2})$.
\begin{eqnarray*}
Y^{ijk} &=&  [X^{i},[X^{j},P^{k}]] = [X^{i},\,\ii\,(1-A\,|\vec{P}|-A^{2}\,|\vec{P}|^{2})\delta^{jk}-\ii A\,\frac{P^{j}P^{k}}{|\vec{P}|}\,(1+A\,|\vec{P}|)] \\
        &=&  -\ii A\,[X^{i},|\vec{P}|]\,\delta^{jk} - 2\ii A^{2}\,|\vec{P}|\,[X^{i},|\vec{P}|]\,\delta^{jk} - \ii A\,[X^{i},P^{j}]\,P^{k} \frac{(1+A\,|\vec{P}|)}{|\vec{P}|} - \ii A\,P^{j}\,[X^{i},P^{k}] \frac{(1+A\,|\vec{P}|)}{|\vec{P}|} \\
        & &+ \ii A\,\frac{P^{j}P^{k}}{|\vec{P}|^{2}}\,[X^{i},|\vec{P}|]\,(1+A\,|\vec{P}|) - \ii A^{2}\,\frac{P^{j}P^{k}}{|\vec{P}|}\,[X^{i},|\vec{P}|] \;.
\end{eqnarray*}
We need the relation
\begin{equation}\label{XiP}
[X^{i},|\vec{P}|]=[X^{i},(P^{j}\,P_{j})^{1/2}]=\frac{P_{j}[X^{i},P^{j}]}{(P^{j}\,P_{j})^{1/2}}=\frac{P_{j}}{|\vec{P}|}\,[X^{i},P^{j}] \;.
\end{equation}
By using \eqref{XiP} and \eqref{commutators_hypervariables} in the last expression of $Y^{ijk}$, we get, up to order $O(A^{2})$,
{\small\begin{eqnarray*}
Y^{ijk} &=&  -\ii A\,\frac{P_{l}}{|\vec{P}|}\,[X^{i},P^{l}]\,\delta^{jk} - 2\ii A^{2}\,P_{l}\,[X^{i},P^{l}]\,\delta^{jk} - \ii A\,[X^{i},P^{j}]\,P^{k} \frac{(1+A\,|\vec{P}|)}{|\vec{P}|}\\ & & - \ii A\,P^{j}\,[X^{i},P^{k}] \frac{(1+A\,|\vec{P}|)}{|\vec{P}|}
        + \ii A\,\frac{P^{j}P^{k}P_{l}}{|\vec{P}|^{3}}\,[X^{i},P^{l}]\,(1+A\,|\vec{P}|) - \ii A^{2}\,\frac{P^{j}P^{k}P_{l}}{|\vec{P}|^{2}}\,[X^{i},P^{l}] \\
       &=&   A\,\frac{P_{l}}{|\vec{P}|}\,\left((1-A\,|\vec{P}|)\delta^{il}-A\,\frac{P^{i}P^{l}}{|\vec{P}|}\,\right)\,\delta^{jk} + 2 A^{2}\,P_{l}\,\delta^{il}\,\delta^{jk} + A\,\left((1-A\,|\vec{P}|)\delta^{ij}-A\,\frac{P^{i}P^{j}}{|\vec{P}|}\,\right)\,P^{k}\, \frac{(1+A\,|\vec{P}|)}{|\vec{P}|}\\ & &  + A\,P^{j}\,\left((1-A\,|\vec{P}|)\delta^{ik}-A\,\frac{P^{i}P^{k}}{|\vec{P}|}\,\right)\, \frac{(1+A\,|\vec{P}|)}{|\vec{P}|}
        - A\,\frac{P^{j}P^{k}P_{l}}{|\vec{P}|^{3}}\left((1-A\,|\vec{P}|)\delta^{il}-A\,\frac{P^{i}P^{l}}{|\vec{P}|}\,\right)  \\
        &=& A\,\frac{P^{i}}{|\vec{P}|}\,\delta^{jk}-\,A^{2}\,P^{i}\,\delta^{jk}-A^{2}\,P^{i}\,\delta^{jk} + 2 A^{2}\,P^{i}\,\delta^{jk} + A\,\frac{P^{k}}{|\vec{P}|}\,\delta^{ij}-A^{2}\,\frac{P^{i}P^{j}P^{k}}{|\vec{P}|^{2}} \\
        & & + A\,\frac{P^{j}}{|\vec{P}|}\,\delta^{ik}-A^{2}\,\frac{P^{j}P^{i}P^{k}}{|\vec{P}|^{2}}
        - A\,\frac{P^{j}P^{k}P^{i}}{|\vec{P}|^{3}} + A^{2}\frac{P^{j}P^{k}P^{i}}{|\vec{P}|^{2}} + A^{2}\,\frac{P^{j}P^{k}P^{i}}{|\vec{P}|^{2}} \\
        &=& A\,\frac{P^{i}}{|\vec{P}|}\,\delta^{jk} + A\,\frac{P^{k}}{|\vec{P}|}\,\delta^{ij} + A\,\frac{P^{j}}{|\vec{P}|}\,\delta^{ik}-A\,\frac{P^{j}P^{k}P^{i}}{|\vec{P}|^{3}}  \;.
\end{eqnarray*}}
We observe that $Y^{ijk}$ is totally symmetric in the three indices at order $\mbox{O}(A^{2})$. Therefore,
\begin{equation*}
[P^{k},[X^{i},X^{j}]] = - Y^{ijk} + Y^{jik} = 0 \;,
\end{equation*}
up to order $\mbox{O}(A^{2})$. The most natural choice for our algebra is then \eqref{Hyper_commutative_GUP}
\begin{equation*}
  [X^{i} , P^{j}] = \ii\left[ \delta^{ij} - A\, |\vec{P}|\,\left(  \delta^{ij} + \frac{P^{i}P^{j}}{|\vec{P}|^{2}} \right)
  - A^{2} \,|\vec{P}|^{2}\,\left( \delta^{ij} + \frac{P^{i}P^{j}}{|\vec{P}|^{2}} \right) \right] , \ \ \ [X^{i} , X^{j}] =  0  , \ \ \  [P^{i} , P^{j}] = 0  \,.
\end{equation*}

\section{Canonical hypervariables from canonical supervariables: More details}
\label{canonical_hyper_from_canonical_super_appendix}

Let us write first some useful identities. As $P^{i}=P^{i}_{0}(1-A P_{0})$,
\begin{equation}
|\vec{P}|\equiv(P^{i}P_{i})^{1/2}=(P_{0}^{i}P_{0i})^{1/2}(1-A\,|\vec{P_{0}}|)=|\vec{P_{0}}|(1-A\,|\vec{P_{0}}|) \;.
\end{equation}
Also,
\begin{equation}\label{property1}
[X_{0}^{i},|\vec{P_{0}}|]=[X_{0}^{i},(P_{0}^{j}P_{0j})^{1/2}]=\frac{[X_{0}^{i},P_{0}^{k}]P_{0\,k}}{(P_{0}^{j}P_{0j})^{1/2}}= \ii\frac{P^{i}_{0}}{|\vec{P_{0}}|} \;,
\end{equation}
where in the last equation we used that $(\vec{X}_{0},\vec{P}_{0})$ are canonical supervariables.

Now, we want to express the hypercoordinates $X^{i}$ in terms of the supervariables, $X^{i}=X^{i}(\vec{X_{0}},\vec{P}_{0})$. Taking the ansatz
\begin{equation}\label{ansatz}
X^{i} = X^{i}_{0} + F\,X^{i}_{0} + G\,A^{2}\,(X^j_{0}P_{0\,j})\,P^{i}_{0} \;,
\end{equation}
where $F=F(A\,|\vec{P_{0}}|)$ and $G=G(A\,|\vec{P_{0}}|)$ are two arbitrary dimensionless functions of $A\,|\vec{P_{0}}|$ (which is also dimensionless). Then,
\begin{equation*}
\begin{split}
[X^{i},P^{j}] &= [X^{i}_{0} + F\,X^{i}_{0} + G\,A^{2}\,(X^k_{0}P_{0\,k})\,P^{i}_{0}, P^{j}_{0} - A \,|\vec{P_{0}}|\, P^{j}_{0}] \\
              &= [X^{i}_{0},P^{j}_{0}] - A\,[X^{i}_{0},|\vec{P_{0}}|]P^{j}_{0} - A\,|\vec{P_{0}}|[X^{i}_{0},P^{j}_{0}] + F[X^{i}_{0},P^{j}_{0}] - F\,A\,[X^{i}_{0},|\vec{P_{0}}|]P^{j}_{0} - F\,A\,|\vec{P_{0}}|[X^{i}_{0},P^{j}_{0}] \\
              &+ G\,A^{2}\,P^{k}_{0}P^{i}_{0}[X^{k}_{0},P^{j}_{0}] - G\,A^{3}\,P^{k}_{0}P^{i}_{0}P^{j}_{0}[X^{k}_{0},|\vec{P_{0}}|] - G\,A^{3}\,P^{k}_{0}\,P^{i}_{0}\,|\vec{P_{0}}|[X^{k}_{0},P^{j}_{0}] \;,
\end{split}
\end{equation*}
where the sum is implicit when repeated indices appear. By using the canonical brackets of supervariables, $[X^{i}_{0},P^{j}_{0}]= \ii\delta^{ij}$, and (\ref{property1}), we have
\begin{equation*}
\begin{split}
[X^{i},P^{j}] &= \ii \delta^{ij} - \ii A\,\frac{P^{i}_{0}}{P_{0}}P^{j}_{0} - \ii A\,|\vec{P_{0}}|\,\delta^{ij} + \ii F\,\delta^{ij} - \ii F\,A\frac{P^{i}_{0}}{|\vec{P_{0}}|}P^{j}_{0} - \ii F\,A\,|\vec{P_{0}}|\,\delta^{ij}  \\
              &+ \ii G\,A^{2}P^{k}_{0}P^{i}_{0}\delta^{kj} - \ii G\,A^{3}P^{k}_{0}P^{i}_{0}P^{j}_{0}\frac{P^{k}_{0}}{|\vec{P_{0}}|} - \ii G\,A^{3}P^{k}_{0}P^{i}_{0}\,|\vec{P_{0}}|\delta^{kj} \\
              &= \ii(1-A\,|\vec{P_{0}}|)(1+F)\delta^{ij}  - \ii\frac{A\,P^{i}_{0}P^{j}_{0}}{|\vec{P_{0}}|}(1+F-G\,A\,|\vec{P_{0}}|(1-2A\,|\vec{P_{0}}|))  \;.
\end{split}
\end{equation*}

In this case we demand the hypervariables to satisfy the algebra (\ref{commutators2})
\begin{equation*}
[X^{i},P^{j}]= \ii \delta_{ij} \;, \; [X^{i},X^{j}]=0 \;, \; [P^{i},P^{j}]=0 \;.
\end{equation*}

Therefore we have the following algebraic system
\begin{eqnarray*}
(1-A\,|\vec{P_{0}}|)(1+F)&=&1 \\
1+F-G\,A\,|\vec{P_{0}}|(1-2AP_{0})&=&0 \;,
\end{eqnarray*}
whose unique solution is, unless $A\,|\vec{P_{0}}|=1$ or $A\,|\vec{P_{0}}|=1/2$,
\begin{eqnarray}
F&=&\frac{A\,|\vec{P_{0}}|}{1-A\,|\vec{P_{0}}|} \\
G&=&\frac{1}{A\,P_{0}(1-A\,|\vec{P_{0}}|)(1-2A\,|\vec{P_{0}}|)} \;.
\end{eqnarray}
Inserting this solution into (\ref{ansatz}) we end up with (\ref{first_relation}).

\section{GUP hypervariables from GUP supervariables: More details}
\label{GUP_hyper_from_GUP_super_appendix}

Here we verify that (\ref{hyper_GUP}) fulfills the Jacobi identity (\ref{Jacobi_X_X_P}). As $[X^{i},P^{j}]= \ii \mathcal{F}^{i\,j}$ then
\begin{equation*}
Y^{i\,j\,k} \equiv [X^{i},[X^{j},P^{k}]] = - \tensor{\mathcal{F}}{^{j\,k}_{\,,l}} \, \mathcal{F}^{i\,l} \;,
\end{equation*}
where ${\mathcal{F}^{ij}}_{, k} = \partial \mathcal{F}^{ij} / \partial P^{k}$. With this definition of $Y^{i\,j\,k}$, we can check that
\begin{align*}
Y^{1\,2\,1} &= A\,\left(\frac{P^{2}}{|\vec{P}|}\right)^{3} + 2 \ell\,A\, \frac{P^{1}\,P^{2}}{|\vec{P}|} + \frac{\ell^{2}}{4}\,P^{2} + \mbox{O}(\ell^{3}) \;, \\
Y^{2\,1\,1} &= A\,\left(\frac{P^{2}}{|\vec{P}|}\right)^{3} + 2 \ell\,A\, \frac{P^{1}\,P^{2}}{|\vec{P}|} + \frac{\ell^{2}}{4}\,P^{2} + \mbox{O}(\ell^{3}) \;, \\
Y^{1\,2\,2} &=  A\,\left(\frac{P^{1}}{|\vec{P}|}\right)^{3} + \ell\,\left(A\,\frac{(P^{1})^{2}}{2|\vec{P}|} + \frac{5(P^{2})^{2}}{2|\vec{P}|} - \frac{1}{2}\right)  + \frac{\ell^{1}}{4}\,P^{1} + \mbox{O}(\ell^{3}) \;, \\
Y^{1\,2\,1} &= A\,\left(\frac{P^{1}}{|\vec{P}|}\right)^{3} + \ell\,\left(A\,\frac{(P^{1})^{2}}{2|\vec{P}|} + \frac{5(P^{2})^{2}}{2|\vec{P}|} - \frac{1}{2}\right)  + \frac{\ell^{1}}{4}\,P^{1} + \mbox{O}(\ell^{3}) \;.
\end{align*}
These equalities up to order $\mbox{O}(\ell^{2})$ are enough to show by exhaustivity (as $i,j$ and $k$ can take only the values $1$ and $2$) that,
\begin{equation*}
- [X^{i},[X^{j},P^{k}]] +  [X^{j},[X^{i},P^{k}]] = 0 \;.
\end{equation*}
Then, the Jacobi identity is fulfilled if $[P^{k},[X^{i},X^{j}]]=0$. This is of course true for $[X^{i},X^{j}]=0$, as was taken in (\ref{hyper_GUP}).

\bibliographystyle{apsrev4-2}
\bibliography{three_layers_GUP_paper_biblio}

\end{document}